\def\hetetwo{{\it HETE-2}\ }

\def\bsax{{\it Beppo}SAX\ }

\def\thetajet{\theta_{\rm jet}}
\def\thetav{\theta_{\rm view}}
\def\ojet{\Omega_{\rm jet}}
\def\ojetmin{\Omega_{\rm jet}^{\rm min}}
\def\ojetmax{\Omega_{\rm jet}^{\rm max}}
\def\oview{\Omega_{\rm view}}

\def\oindex{\delta}

\def\ejet{E_{\rm jet}}
\def\eiso{E_{\rm iso}}
\def\egamma{E_{\gamma}}
\def\ep{E_{\rm peak}}
\def\eop{E^{\rm obs}_{\rm peak}}
\def\se{S_E}
\def\sn{S_N}
\def\fe{F^{\rm P}_E}
\def\fn{F^{\rm P}_N}
\def\liso{L_{\rm iso}}

\def\lgamma{L_{\gamma}}
\def\lx{L_{\rm X,iso}}

\def\eradio{E_{\rm radio}}

\documentclass[12pt,preprint]{aastex}

\shorttitle{Unified Jet Model of XRFs, X-Ray-Rich GRBs, and GRBs}
\shortauthors{Lamb, Donaghy, and Graziani}

\begin{document}

\newcommand{\beq}{\begin{equation}}
\newcommand{\eeq}{\end{equation}}

\title{A Unified Jet Model of X-Ray Flashes, X-Ray-Rich GRBs, and GRBs:
I. Power-Law-Shaped Universal and Top-Hat-Shaped Variable Opening-Angle
Jet Models}

\author{D. Q. Lamb, T. Q. Donaghy, and C. Graziani}
\email{d-lamb@uchicago.edu, quinn@oddjob.uchicago.edu, carlo@oddjob.uchicago.edu}
\affil{Department of Astronomy and Astrophysics, University of
Chicago, 5640 S. Ellis Avenue, Chicago, IL 60637}

\begin{abstract} 
HETE-2 has provided strong evidence that the properties of X-Ray
Flashes (XRFs), X-ray-rich GRBs, and GRBs form a continuum, and
therefore that these three kinds of bursts are the same phenomenon.  A
key feature found by HETE-2 is that the density of bursts is roughly
constant per logarithmic interval in burst fluence $\se$ and observed
spectral peak energy $\eop$, and in isotropic-equivalent energy $\eiso$
and spectral peak energy $\ep$ in the rest frame of the burst.  In this
paper, we explore a unified jet model of all three kinds of bursts,
using population synthesis simulations of the bursts and detailed
modeling of the instruments that detect them.  We show that both a
variable jet opening-angle model in which the emissivity is a constant
independent of the angle relative to the jet axis and a universal jet
model in which the emissivity is a power-law function of the angle
relative to the jet axis can explain the observed properties of GRBs
reasonably well.  However, if one tries to account for the properties
of XRFs, X-ray-rich GRBs, and GRBs in a unified picture, the extra
degree of freedom available in the variable jet opening-angle model
enables it to explain the observations reasonably well while the
power-law universal jet model cannot.  The variable jet opening-angle
model of XRFs, X-ray-rich GRBs, and GRBs implies that the energy
$\egamma$ radiated in gamma rays is $\sim$ 100 times less than has been
thought.  The model also implies that most GRBs have very small jet
opening angles ($\sim$ half a degree).  This suggests that magnetic
fields may play an important role in GRB jets.   It also implies that
there are $\sim 10^4 -10^5$ more bursts with very small jet opening
angles for every burst that is observable.  If this is the case, the
rate of GRBs could be comparable to the rate of Type Ic core collapse
supernovae.  These results show that XRFs may provide unique
information about the structure of GRB jets, the rate of GRBs, and the
nature of Type Ic supernovae.
\end{abstract}

\keywords{gamma rays: bursts --- supernovae: general --- ISM: jets and outflows --- shock waves}

\section{Introduction}

One-third of all HETE-2--localized bursts are ``X-ray-rich'' GRBs and an
additional one-third are XRFs \footnote{We define ``X-ray-rich'' GRBs
and XRFs as those events for which $\log [S_X(2-30~{\rm kev}) /
S_\gamma(30-400~{\rm kev})] > -0.5$ and 0.0, respectively.}
\citep{sakamoto2003b}.  The latter have received increasing attention
in the past several years \citep{heise2000,kippen2002}, but their
nature remains largely unknown.

XRFs have $t_{90}$ durations between 10 and 200 sec and their sky
distribution is consistent with isotropy \citep{heise2000}.  In these
respects, XRFs are similar to GRBs.  A joint analysis of WFC/BATSE
spectral data showed that the low-energy and high-energy photon indices
of XRFs are $-1$ and $\sim -2.5$, respectively, which are similar to
those of GRBs, but that the XRFs have spectral peak energies $E_{\rm
peak}^{\rm obs}$ that are much lower than those of GRBs
\citep{kippen2002}. The only difference between XRFs and GRBs therefore
appears to be that XRFs have lower $E_{\rm peak}^{\rm obs}$ values.  It
has therefore been suggested that XRFs might represent an extension of
the GRB population to bursts with low peak energies, and that the
distinction bewteen XRFs and GRBs is driven by instrumental
considerations rather than by any sharp intrinsic difference between
the two kinds of bursts \citep{kippen2002,barraud2003,sakamoto2003b}.

A number of theoretical models have been proposed to explain XRFs. 
\cite{yamazaki2002,yamazaki2003} have proposed that XRFs are the result
of a highly collimated GRB jet viewed well off the axis of the jet.  In
this model, the low values of $\ep$ and $\eiso$ (and therefore for
$\eop$ and $\se$) seen in XRFs is the result of relativistic beaming. 
However, it is not clear that such a model can produce roughly equal
numbers of XRFs, XRRs, and GRBs, and still satisfy the observed
relation between $\eiso$ and $\ep$ \citep{amati2002,lamb2003}.

According to \citet{meszaros2002} and \citet{wzhang2004}, X-ray
(20-100 keV) photons are produced effectively by the hot cocoon
surrounding the GRB jet as it breaks out, and could produce XRF-like
events if viewed well off the axis of the jet.  However, it is not
clear that such a model would produce roughly equal numbers of XRFs,
XRRs, and GRBs, or the nonthermal spectra exhibited by XRFs.
 
The ``dirty fireball'' model of XRFs posits that baryonic material is
entrained in the GRB jet, resulting in a bulk Lorentz factor $\Gamma$
$\ll$  300 \citep{dermer1999,huang2002,dermer2003}.  At the opposite
extreme, GRB jets in which the bulk Lorentz factor $\Gamma$ $\gg$ 300
and the contrast between the bulk Lorentz factors of the colliding
relativistic shells are small can also produce XRF-like events
\citep{mochkovitch2003}.

In this paper, we explore a unified jet picture of XRFs, X-ray-rich
GRBs, and GRBs, motivated by HETE-2 observations of the three kinds of
bursts.  We consider two different phenomenological jet models: a
variable jet opening-angle model in which the emissivity is a constant
independent of the angle relative to the jet axis and a universal jet
model in which the emissivity is a power-law function of the angle
relative to the jet axis.  We show that the variable jet opening-angle
model can account for the observed properties of all three kinds of
bursts.   In contrast, we find that, although the power-law universal
jet model can can account reasonably well for the observed properties
of GRBs, it cannot easily be extended to account for the observed
properties of XRFs and X-ray-rich GRBs.

This paper is organized as follows.  In \S 2, we summarize the results
of HETE-2 observations of XRFs, X-ray-rich GRBs, and GRBs.  In \S 3, we
define the variable jet opening angle model and the power-law universal
jet model.  In \S 4, we describe our simulations, detailing how we
model the bursts themselves, propagate the bursts to the Earth, and
model the instruments that detect them.  In \S 5, we compare our
results with observations, and in \S 6, we discuss their implications. 
In \S 7 we present our conclusions.  Preliminary results were reported
in Lamb, Donaghy \& Graziani (2004a,b,c).

\section{Observations of XRFs and GRBs}

\subsection{HETE-2 Results}

\begin{figure}[t]
\includegraphics[width=4.7truein,clip=,angle=270]{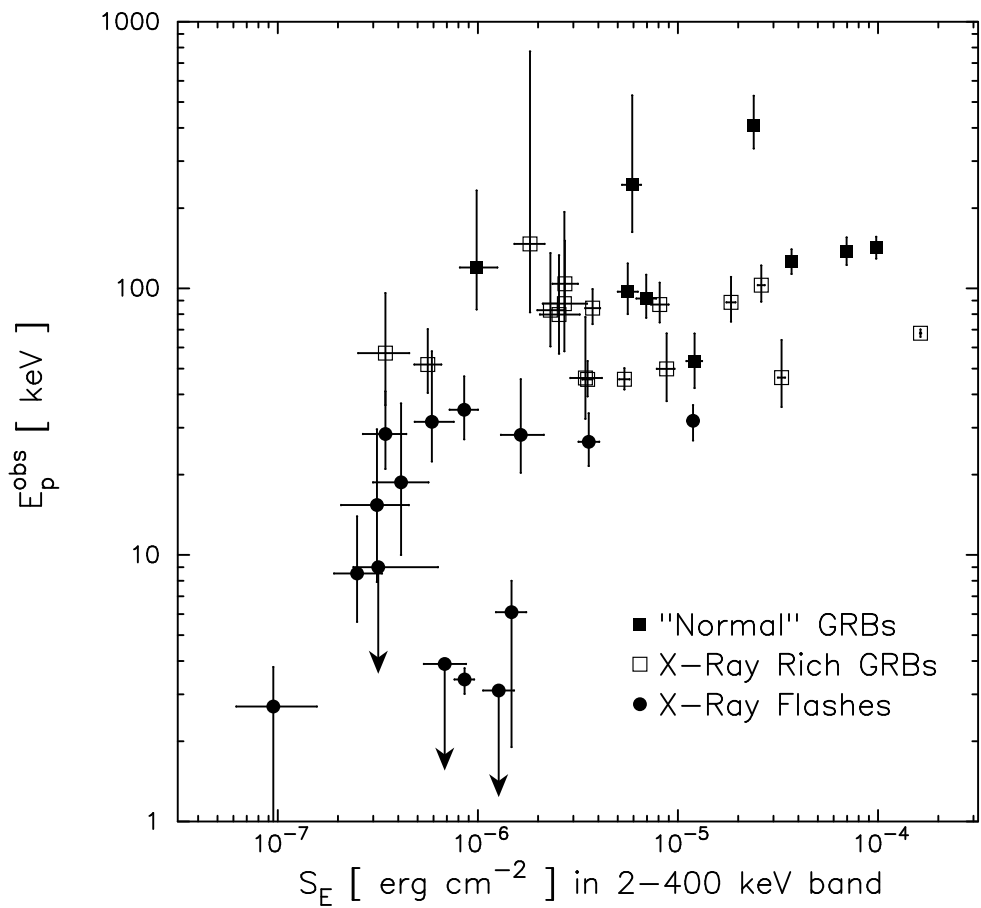}
\caption{Distribution of HETE-2 bursts in the [$S(2-400 {\rm keV}),
E^{\rm obs}_{\rm peak}$]-plane, showing XRFs (filled circles),
X-ray-rich GRBs (open boxes), and GRBs (filled boxes).  This figure
shows that XRFs and X-ray-rich GRBs comprise about 2/3 of the bursts
observed by \hetetwo, and that the properties of the three kinds of
bursts form a continuum.  All error bars are 90\% confidence level. 
From \cite{sakamoto2003b}.} 
\label{hete_Epobs_flu_by_class}
\end{figure}

Clarifying the nature of XRFs and their connection to GRBs could
provide a breakthrough in our understanding of the prompt emission of
GRBs, and of the structure of XRF and GRB jets.  Analyzing 45 bursts
seen by the FREGATE \citep{atteia2003} and/or the WXM \citep{kawai2003}
instruments on HETE-2 \citep{ricker2003}, \cite{sakamoto2003b} find
that XRFs, X-ray-rich GRBs, and GRBs form a continuum in the
[$S_E(2-400~{\rm kev}),\eop$]-plane (see Figure
\ref{hete_Epobs_flu_by_class}), where $S_E(2-400~{\rm kev})$ is the
fluence of the burst in the 2-400 keV energy band and $\eop$ is the
energy of the observed peak of the burst spectrum in $\nu F_\nu$. 

Furthermore, \cite{lamb2003} have placed 9 HETE-2 GRBs with known
redshifts and 2 XRFs with known redshifts or strong redshift
constraints in the ($\eiso,\ep$)-plane (see Figure \ref{amati_plots}). 
Here $\eiso$ is the isotropic-equivalent burst energy and $E_{\rm
peak}$ is the energy of the peak in the burst $\nu F_\nu$ spectrum,
measured in the source frame.   We define $\eiso$ to be the energy
emitted in the source-frame passband from $1-10000$ keV.  This
definition is a suitable bolometric quantity for both GRBs and XRFs,
and is the same definition of $\eiso$ used by \cite{amati2002}.  The
HETE-2 results confirm the correlation between $\eiso$ and $\ep$ found
by \cite{lloyd-ronning2000} for BATSE bursts and the relation between
these two quantities found for \bsax bursts by \cite{amati2002} for
\bsax bursts with known redshifts, and extend it down in $\eiso$ by a
factor of $\sim$ 300.  The fact that XRF 020903 \citep{sakamoto2003a},
the softest burst localized by HETE-2 to date, and XRF 030723
\citep{prigozhin2003}, lie squarely on this relation \citep{lamb2003}
is evidence that the relation between $\eiso$ and $\ep$ extends down in
$\eiso$ by a factor of $\sim$ 300 and applies to XRFs and X-ray-rich
GRBs as well as to GRBs.  However, additional redshift determinations
are clearly needed for XRFs with 1 keV $< \ep < 30$ keV in order to
confirm this.

\cite{lamb2003} show that, using HETE-2 and \bsax GRBs with known
redshifts and XRFs with known redshifts or strong redshift
constraints, there is also a relation between the isotropic-equivalent
burst luminosity $\liso$ and $\ep$ that extends over five decades in
$\liso$, and (as must then be the case) between $\eiso$ and
$\liso$ that extends over five decades in both (see Figure
\ref{amati_plots}).  \cite{yonutoku2004} have confirmed the relation
between $\liso$ and $\ep$ for GRBs, while \cite{liang2004} have
shown that this relation holds within GRBs.

Thus the HETE-2 results that show that the properties of XRFs,
X-ray-rich GRBs, and GRBs form a continuum in the [$S_E(2-400~{\rm
kev}),\eop$]-plane and that the relation between $\eiso$ and $\ep$
extends to XRFs and X-ray-rich GRBs.  A key feature of the
distributions of bursts in these two planes is that the density of
bursts is roughly constant along these relations, implying roughly
equal numbers of bursts per logarithmic interval in $S_E$, $\eop$,
$\eiso$ and $\ep$.  These results, when combined with the earlier
results described above, strongly suggest that all three kinds of
bursts are the same phenomenon.  It is this possibility that
motivates us to seek a unified jet model of XRFs, X-ray-rich GRBs, 
and GRBs.

\subsection{Evidence That Most GRBs Have a ``Standard'' Energy}

\cite{frail2001} and \cite{panaitescu2001} [see also \cite{bloom2003}]
find that most GRBs have a ``standard'' energy.  That is, most GRBs
have the same radiated energy, $\egamma = 1.3 \times 10^{51}$ ergs, to
within a factor $\sim$ 2-3, if their isotropic equivalent energy is
corrected for the jet opening angle $\thetajet$ inferred from the jet
break time.  This is illustrated in Figure \ref{bloom_fig1}, which
shows the distribution of total radiated energies in gamma-rays
$\egamma$ for 24 GRBs, after taking into account the jet opening angle
inferred from the jet break time \citep{bloom2003}.

Pursuing this picture further, we show in Figure \ref{ojet_corrs} the
distribution of $\eiso$, $\liso$, and $\ep$ as a function of
$2\pi/\ojet$ for the HETE-2 and \bsax GRBs with known redshifts. 
Figure \ref{ojet_corrs} shows that all three quantities are strongly
correlated with $\ojet$.  The correlation between $\eiso$ and $\ojet$
is implied by the fact that most GRBs have a standard energy
\citep{frail2001,panaitescu2001}.  The correlation between $\liso$ and
$\ojet$ is implied by the fact that most GRBs have a standard
energy and the correlation between $\eiso$ and $\liso$
\citep{lamb2003}.  The correlation between $\ep$ and $\ojet$ is implied
by the fact that most GRBs have a standard energy, and the
correlation between $\eiso$  and $\ep$ found by
\cite{lloyd-ronning2000} for BATSE bursts without redshifts and the
tight relation between $\eiso$ and $\ep$ found by\cite{amati2002} for
\bsax bursts with known redshifts.  Figure \ref{ojet_corrs}
demonstrates these three correlations directly.

The strength of the correlations of all three quantities with $\ojet$
lends additional support to a picture in which most GRB have a
standard energy and the observed ranges of $\sim 10^5$ in $\eiso$ and
$\liso$ are due either to differences in the jet opening angle
$\thetajet$ or to differences in the viewing angle $\thetav$ of the
observer with respect to the axis of the jet.  We pursue both of
these possibilities below.

\section{Jet Models of GRBs}

Two phenomenological models of GRB jets have received widespread
attention:

\begin{itemize}

\item 
The universal jet model (see the left-hand panel of Figure
\ref{two_jet_schematics}).  In this model, all GRBs produce jets with
the same structure
\citep{rossi2002,zhang2002,meszaros2002,wzhang2004,perna2003,zhang2004}. 
The energy $\eiso$ and luminosity $\liso$ are assumed to decrease as
the viewing angle $\thetav$ increases.  The wide range of observed
values of $\eiso$ is then attributed to differences in the viewing
angle $\thetav$.  In order to recover the standard energy result
\citep{frail2001,panaitescu2001,bloom2003} over a wide range in viewing
angles, $\eiso (\thetav) \propto \thetav^{-2}$ is required
\citep{rossi2002,zhang2002}.
\bigskip

\item 
The variable jet opening-angle model (see the right-hand panel of
Figure \ref{two_jet_schematics}).  In this model, GRB jets have a wide
range of jet opening angles $\thetajet$ \citep{frail2001}.  For
$\thetav < \thetajet$, $\eiso (\thetav) \approx$ constant, while for
$\thetav > \thetajet$, $\eiso (\thetav) = 0$.  The wide range of
observed values of $\eiso$ is then attributed to differences in the jet
opening angle $\thetajet$.  This is the model that \cite{frail2001} and
\cite{bloom2003} assume in deriving a standard energy for most bursts.

\end{itemize}

As described in the previous section, there is evidence that the
relation between $\eiso$ and $\ep$ extends over at least five decades
in $\eiso$, and appears to hold for XRFs and X-ray-rich GRBs, as well
as for GRBs \citep{lamb2003}; most bursts appear to have a standard
energy \citep{frail2001,panaitescu2001,bloom2003}; and there are
correlations among $\eiso$, $\liso$, and $\ep$, and between these
quantities $\ojet$ \citep{frail2001,bloom2003,lamb2003}.  Motivated by
these results, we make three key assumptions in exploring a unified
jet picture of all three kinds of bursts:

\begin{enumerate}

\item
We assume that most XRFs, X-ray-rich GRBs, and GRBs have a standard
energy $\egamma$ with a modest scatter.

\item
We assume that, for most GRBs, $\eiso$ and $\ep$ obey the relation
\citep{lloyd-ronning2000,amati2002,lamb2003}:
\beq
E_{\rm peak} \propto (\eiso)^{1/2} \; ,
\eeq
with a modest scatter, and that this relation holds for XRFs and
X-ray-rich GRBs, as well as for GRBs.

\item
We assume that the observed ranges of $\sim 10^5$ in $\eiso$ and
$\liso$ are due either to differences in the jet opening angle
$\thetajet$ (in the variable jet opening-angle model) or to differences
in the viewing angle $\thetav$ of the observer with respect to the axis
of the jet (in the universal jet model). 

\end{enumerate}

\section{Simulations of Observed Gamma-Ray Bursts}

\subsection{Overview of the Simulations}

We begin by giving an overview of our population synthesis simulations
of observed GRBs before describing the simulations in mathematical
detail.  Our overall approach is to simulate the GRBs that are observed
by different instruments by (1) modeling the bursts in the source
frame; (2) propagating the bursts from the source frame to us, using
the cosmology that we have adopted; and (3) determining which bursts
are observed and the properties of these bursts by modeling the
instruments that observe them.  

This logical sequence is evident in Figure \ref{flowchart}, which shows
a flowchart of the calculations involved in our simulations of bursts
in the variable jet opening-angle model.   For each simulated burst we
obtain a redshift $z$ and a jet opening solid angle $\ojet$ by drawing
from specific distributions.  In addition, we introduce three
lognormal smearing functions to generate a timescale $T$, a jet energy
$\egamma$ and a coefficient for the $\eiso-\ep$ relation ($C$).  Using
these five quantities, we calculate various rest-frame quantities
($\eiso$, $\liso$, $\ep$, etc.).  Finally, we construct a Band spectrum
for each burst and transform it into the observer frame, which allows
us to calculate fluences and peak fluxes, and to determine if the burst
would be detected by various experiments.

Astronomical observations usually impose strong observational selection
effects on the population of objects being observed.   Consequently,
the most rigorous approach to comparing models to data, and finding the
best-fit parameters for these models, is to specify the models being
compared, independent of any observations.  This avoids the pitfall of
circularity, in which the posited models are already distorted by
strong observational selection effects.  In practice, this approach is
difficult to carry out, particularly when our understanding of the
phenomenon of interest is quite limited, as is currently the case for
GRB jets.  

We therefore adopt an intermediate approach in this paper.  We use
those properties of GRBs that we have reason to believe are unlikely
to be strongly affected by observational selection effects as a guide
in specifying the models that we consider.  We then extend the
predictions of these models to regimes in which the observational
selection effects are strong by modeling these effects in detail.  We
are then able to compare the predictions of the models with
observations in the regimes where we believe observational selection
effects are unlikely to be important and in the regimes where we know
that observational selection effects are important.

\subsection{GRB Rest-Frame Quantities}

\subsubsection{Variable Jet Opening-Angle Model}

In this paper we investigate a variable jet opening-angle model in
which the emissivity is a constant independent of the angle relative to
the jet axis and the distribution of jet opening angles is a power law.
In a subsequent paper, we investigate variable jet opening-angle models
in which the emissivity is a constant independent of the angle relative
to the jet axis and the distribution of jet opening angles is a
Gaussian, and in which the emissivity is a Gaussian function of the
angle relative to the jet axis and the distribution of jet opening
angles is a power law \citep{donaghy2004a}.

We assume that the emission from the jet is visible only when $\thetav
< \thetajet$.  In reality, emission from the jet may be seen when the
observer is outside the opening angle of the jet, due to relativistic
beaming effects.  However, the angular width of the annulus within
which the jet is visible (i.e., has a flux above some minimum
observable flux) is small.  If the opening angle of the jet is large
(as is posited to be the case for XRFs in the variable jet opening
angle model), the relative number of bursts that will be detectable
because of relativistic beaming is therefore also small
\citep{donaghy2004c}.  If the opening angle of the jet is small (as is
posited to be the case for GRBs in the variable jet opening angle
model), the bulk $\Gamma$ in the jet may be large and the flux due to
relativistic beaming that is seen by an observer outside the opening
angle will then drop off precipitously.  The relative number of bursts
that will be detectable because of relativistic beaming is therefore
again small \citep{donaghy2004c}.

The distribution in jet opening solid angle $\ojet$ then generates our
GRB luminosity function; here we are primarily interested in a
power-law distribution.  We define the fraction of the sky subtended
by the GRB jet to be
\beq
f_{\rm jet} = \frac{\ojet}{2\pi} = 
	1 - \cos \theta _{\rm jet} \; .
\eeq
We define the true distribution of opening angles to be 
\beq
P_{\rm true}(\ojet) d\ojet 
= {\rm const} \times (\ojet)^{-\oindex} d\ojet
\eeq
over a range ($\ojetmin,\ojetmax$).
We define the observed distribution of opening angles to be
\beq
P_{\rm obs}(\ojet) d\ojet 
= {\rm const} \times (\ojet)^{-\oindex_{\rm sim}} d\ojet
\propto f_{\rm jet} ^{-\oindex_{\rm sim}}.
\eeq
Since we can observe only those bursts whose jets are oriented toward
the Earth, the distribution of opening angles of observable bursts is
related to the true distribution of opening angles by 
\beq
P_{\rm obs}(\ojet) = f_{\rm jet} P_{\rm true}(\ojet) 
\propto (\ojet)^{(1 -\oindex)}
\eeq
We thus simulate bursts using the power-law index $\oindex_{\rm sim}$
from which the true power-law index can be found using the relation
$\oindex = 1 + \oindex_{\rm sim}$.

The isotropic-equivalent emitted energy $\eiso$ is then given by
\beq
\eiso = \frac{\egamma}{f_{\rm jet}} = \frac{\egamma}{(\ojet/2\pi)},
\eeq
where $\egamma$ is the total radiated energy of the burst.  Using a
full maximum likelihood approach, we reproduce the parameters of the
lognormal distribution derived by \cite{bloom2003}, using their sample
of GRBs with observed jet break times (see Figure
\ref{gaussian_fit_plots}).  We find no evidence for any correlation of
$\egamma$ with redshift (see again Figure \ref{gaussian_fit_plots}).  We
therefore draw values for $\egamma$ randomly from the narrow lognormal
distribution defined by
\beq
G(\egamma) d\log\egamma = \exp \left( 
	\frac{-(\log\egamma - \log \egamma^{0})^{2}}{2\sigma_{E}^{2}} 
	\right) d\log\egamma \; ,
\eeq
where $\log\egamma^{0} ({\rm erg}) = 51.070$  and $\log\sigma_{E} =
0.35$ (see also Table \ref{gauss_table}).

Our simulations thus use a value $\egamma^{0} = 1.17 \times
10^{51}\ {\rm ergs}$, which is fully consistent with the value 
$\egamma^{0} = 1.33 \times 10^{51}\ {\rm ergs}$ found by
\cite{bloom2003}.  However, the Bloom, Frail and Kulkarni sample of
GRBs contained no XRFs.  The values of $\eiso$ for XRFs 020903
\citep{sakamoto2003a} and 030723 \citep{lamb2003} are $\sim 100$
times lower than the value of $\egamma$ derived by \cite{frail2001}
and \cite{bloom2003}.  Thus there is no value of the opening solid
angle $\ojet$ that  can accommodate these values of $\eiso$.  Since
we are pursuing a unified jet model of XRFs, X-ray-rich GRBs, and
GRBs, we must be able to accommodate values of $\eiso$ that are $\sim
100$ times less than the value of $\egamma$ derived by
\cite{frail2001} and \cite{bloom2003}.

We therefore introduce the ability to rescale $\log\egamma^{0}$, the
central value of $\egamma$.  This is equivalent to rescaling the range
of $\ojet$, since only $\eiso$ is observed.  In doing so, we note that
the derivation of $\egamma$ is dependent on the coefficient in front of
the relation between the jet-break time and $\thetajet$, and that the
value of this coefficient is uncertain by a factor 4-5 
\citep{rhoads1999,sari1999}.

This rescaling of $\egamma$ introduces an additional parameter 
$C_{\rm jet}$ into our model:
\beq
\eiso = \frac{\egamma}{C_{\rm jet} \cdot \ojet/2\pi}.
\eeq
XRF 020903, the dimmest burst in our sample, has $\eiso=2.3 \times
10^{49}\ {\rm ergs}$ \citep{sakamoto2003a}.  Accounting for this burst
requires that $C_{\rm jet}$ be at least $57.8$; this choice is
conservative in the sense that it implies that XRF 020903 lies at the
faintest end of the range of possible values of $\eiso$ and has the
maximum possible opening angle of $\ojet=2\pi$.  The brightest burst in
our sample is GRB 990123, which has $\eiso=2.8 \times 10^{54}\ {\rm
ergs}$.  Thus the range of $\eiso$ is at least $\sim 10^5$, and so the
range of $\ojet$ must also be $\sim 10^5$.  Since only $\eiso$ is a
directly observable quantity, the value of $C_{\rm jet}$ is degenerate
with the value of the jet opening solid angle $\ojet$.  Thus GRB 990123
provides a constraint only on $C_{\rm jet} \cdot \Omega_{\rm jet}^{\rm
min}$.

Since we wish our burst simulations to explain the full range of
observed $\eiso$, we require a range of approximately five decades in
$\ojet$ (conservatively, from $2\pi$ to $2\pi \times 10^{-5}\ {\rm
sr}$).  We have then varied $C_{\rm jet}$ to best match the observed
cumulative distributions shown in Figure \ref{fab_four}, as determined
by visual comparison of the observed and predicted cumulative
distributions.  The fiducial model that we use in this paper has a
value of $C_{\rm jet}=95$.  This gives minimum and maximum values of
$\eiso$ of $1.4\times 10^{49}\ {\rm ergs}$ and $1.4\times 10^{54}\ {\rm
ergs}$.  The former value of $\eiso$ implies a jet opening angle
$\thetajet = 67^\circ$ for XRF 020903 (the burst with the smallest
value of $\eiso$ in our sample).  The latter value of $\eiso$ is
slightly smaller than the value of $\eiso$ for GRB 990123 (the burst
with the largest value of $\eiso$ in our sample), but the range of
simulated $\egamma$ values, although narrow, is sufficient to account
for this event and events like it.  We have used the value $C_{\rm
jet}=95$ to rescale the $\ojet$ values reported by \cite{bloom2003}
(see Figure \ref{omega_cuml}); this corresponds to making the
coefficient in the relation between the jet break time and $\thetajet$
a factor of $\sim 10$ smaller, and therefore the value of $\thetajet$ a
factor of $\sim 10$ smaller.

Thus the value of $C_{\rm jet}$ that we adopt in this paper requires
that the value of $\theta_{\rm jet}$ corresponding to a given jet break
time be smaller than the value that is typically assumed by a factor of
about ten; i.e., a factor of two more than the uncertainty stated by
\cite{sari1999}.  We return to this point below in the Discussion
section.

We incorporate the relation between $\eiso$ and $\ep$ found by 
\cite{amati2002} and extended by \cite{lamb2003}, using a second
narrow  lognormal distribution, defined by
\beq
\ep = C \cdot \left( \frac{\eiso}{10^{52}\ {\rm erg}} \right)^{s} \; ,
\eeq
and 
\beq
G(C) d\log C 
= \exp \left(\frac{-(\log C - \log C_{0})^{2}}{2\sigma_{C}^{2}} \right) 
d\log C.
\eeq
We set the power-law index $s = 0.5$. Then, using a full maximum
likelihood approach to fit these equations to the HETE-2 and \bsax GRBs
with known redshifts (see Figures \ref{amati_plots} and
\ref{gaussian_fit_plots}), we find maximum likelihood best-fit
parameters $C_{0} = 90.4$ keV and $\sigma_{C} = 0.70$ (see also Table
\ref{gauss_table}).  Again we find no evidence for any correlation of
$C$ with redshift (see Figure \ref{gaussian_fit_plots}).  We therefore
draw randomly from this Gaussian distribution to choose the value of
$\ep$ corresponding to the value of $\eiso$ for a particular burst.

Finally, we require the timescale that converts the
isotropic-equivalent energy $\eiso$ of a burst to the
isotropic-equivalent peak luminosity $\liso$ of a burst.  Using a full
maximum likelihood approach, we determine this timescale by fitting a
third narrow lognormal distribution, defined by
\beq
G(T) d\log T 
= \exp \left(\frac{-(\log T - \log T_{0})^{2}}{2\sigma_{T}^{2}} \right)
d\log T \; ,
\eeq
to the distribution of the ratio $\eiso$/$\lgamma$ for the HETE-2 and
\bsax bursts with known redshifts (see Figure
\ref{gaussian_fit_plots}).  Thus the timescale $T$ is defined in the
rest  frame of the GRB source.  We find maximum likelihood best-fit
parameters $T_{0} = 3.41$ sec and $\sigma_{T} = 0.33$ (see also Table
\ref{gauss_table}).  Again, we find no evidence for any correlation of
$T$ with redshift (see Figure \ref{gaussian_fit_plots}).  We therefore
draw randomly from this Gaussian distribution and use the formula
$\liso =  \eiso/T$ to convert $\eiso$ to $\liso$, and thus also to
convert burst fluences to peak fluxes.  We note that the sample used
for this fit also contains no XRFs.

\subsubsection{Power-Law Universal Jet Model}

In order to recover the standard energy result in the universal jet
model requires $\eiso \propto \thetav ^{-2}$
\citep{rossi2002,zhang2002,perna2003}.  Therefore in this paper we 
investigate a universal jet model in which the emissivity is a
power-law function of the angle relative to the jet axis.  In a
subsequent paper \citep{donaghy2004a}, we investigate a universal jet
model in which the emissivity is a Gaussian function of the angle
relative to the jet axis \citep{zhang2004}).

The requirement $\eiso \propto \thetav ^{-2} \propto \oview^{-1}$
allows us to simulate the power-law universal jet model by simply
making the substitution $\ojet \to \oview$ in the variable jet
opening-angle simulations.  To see this, compare Equation 6 with
the relation:
\beq
\eiso \propto \frac{1}{\thetav ^{2}} \propto
	\frac{1}{\oview}.
\eeq 
Although the physical interpretations of the two equations are entirely
different, they give the same results.  In addition to this
substitution, we have to specify $\delta_{\rm sim}$ for the power-law
universal jet model.  Since the bursts are randomly oriented with
respect to our line of sight, we draw $\oview$ values from a flat
distribution, $d\oview$, which corresponds to $\delta_{\rm sim}=0$. 
Drawing from this distribution results in very few small $\thetav$
values compared to the very large number of $\thetav$ values near
$\theta_{\rm view, max}$ (the angular extent of the universal jet) or
$90^{\circ}$, whichever is smaller.  Therefore, in this model, most
bursts have $\thetav \sim \theta_{\rm view, max}$ or $90^{\circ}$,
whichever is smaller; and the range of observed $\oview$ values in
logarithmic space is small for a finite sample of bursts.  As a result,
the power-law universal jet model predicts that most of the bursts
arriving at the Earth will have small values of $\eiso$, $\liso$, etc.
\citep{rossi2002,perna2003}.

We also introduce the ability to rescale the central value of $\egamma$
in the power-law universal jet model (see Equation 8).  For this model
we consider two cases: in the first case, we ``pin'' the minimum value
of $\eiso$ (i.e., the value of $\eiso$ corresponding to $\oview=2\pi$)
to the value of $\eiso$ for XRF 020903; in the second case, we pin
the minimum value of $\eiso$ to the value of $\eiso$ for GRB 980326
(the smallest $\eiso$ in our sample of HETE-2 and \bsax bursts with
known redshifts, apart from the XRFs).  In the first case, we derive
$C_{\rm jet} = 58$, and in the second $C_{\rm jet} = 0.24$.  In the
first case, the power-law universal jet model can then generate the
full observed range of $\eiso$ (i.e., both XRFs and GRBs), while in the
second case, it can generate the range of $\eiso$ values corresponding
to GRBs, but not to XRFs or X-ray-rich GRBs.

\subsection{Gamma-Ray Burst Rate as a Function of Redshift}

The observed rate of GRBs per redshift interval $dz$ is given by
\beq
\rho (z) = R_{GRB}(z) \times (1+z)^{-1} \times 4\pi r(z)^{2}
	\frac{dr}{dz} = \left[\rm \frac{num}{dz \cdot year} \right] \; ,
\eeq
where $R_{GRB}(z)$ is the rate of GRBs per comoving volume and $r(z)$
is the comoving distance to the source [see \S 4.3 below for the
precise definition of $r(z)$].  We use the phenomenological
parameterization of the star-formation rate (SFR) as a function of
redshift suggested by \cite{rr2001} to parameterize the GRB rate as a
function of redshift.  In this parameterization, $R_{GRB}$ is given by
\beq
R_{GRB}(z) = R_{0}\, \left(\frac{t(z)}{t(0)}\right)^{P}
	\exp\left[Q\left(1-\frac{t(z)}{t(0)}\right)\right]
	= \left[\rm \frac{num}{year \cdot Gpc^{3}} \right].
\eeq
Here, $t(z)$ is the elapsed coordinate time since the big bang at that
redshift.  In this paper, we adopt the values $P = 1.2$ and $Q = 5.4$,
which provide a good fit to existing data on the star-formation rate
(SFR) as a function of redshift.  The resulting curve of the SFR as a
function of redshift is given in Figure \ref{grb_rate}.  It rises 
rapidly from $z = 0$, peaks at $z \approx 1.5$, and then decreases 
gradually with increasing redshift.  We draw GRB redshifts randomly
from this SFR curve.  

The actual SFR as a function of redshift is uncertain, and the GRB rate
as a function of redshift is even more uncertain.  Several studies have
suggested that the GRB rate may be flat, or may even increase, at high
redshifts \citep{fenimore2000,lloyd-ronning2002,reichart2001b}.  The
particular choice that we have made of the GRB rate as a function of
redshift has little effect on the comparisons with observations that we
carry out in this paper, since all of the bursts that we consider are
at modest redshifts ($z \lesssim 3$).  However, predictions of the
fraction of bursts that lie at very high redshifts ($z > 5$), and
therefore the number of detectable bursts at very high redshifts, are
sensitive to the shape of the GRB rate curve at very high redshifts.

\subsection{Cosmology}

The Rowan-Robinson SFR model depends on a few basic cosmological
parameters, as do the observed peak photon number and energy fluxes and
fluences of the bursts.   In this paper, we adopt the values
$\Omega_{M} = 0.3$, $\Omega_{\Lambda} = 0.7$ and $H_{0} = 65$ km
s$^{-1}$ Mpc$^{-1}$.

The comoving distance to redshift $z$ is defined by 
\beq
\frac{dr}{dz} = \frac{c}{H_{0}} \left[ \Omega_{k} (1+z)^{2} +
	\Omega_{\Lambda} + \Omega_{M} (1+z)^{3} + \Omega_{R} (1+z)^{4}
	\right] ^{-1/2}, 
\eeq
and integrating this equation over $dz$ gives us $r(z)$.

To calculate the time since the big-bang we integrate the following
formula: 
\beq
dt = \frac{da}{a \, H_{0}} \left[ \Omega_{k} a^{-2} +
	\Omega_{\Lambda} + \Omega_{M} a^{-3} + \Omega_{R} a^{-4} \right]
	^{-1/2}, 
\eeq
which yields an expression for $t(z)$.  Here $a = (1+z)^{-1}$.  For our
adopted cosmology, there is an analytic expression for $t(z)$, which
is
\beq
H_{0} t(z) = \frac{2}{3 \Omega_{\Lambda}^{1/2}} \sinh ^{-1}
	\left[ \left( \frac{\Omega_{\Lambda}}{\Omega_{M}} \right)^{1/2}
	(1+z)^{-3/2} \right] \; ,
\eeq
but there is no analytic expression for $r(z)$.

\subsection{Observable Quantities}

In this paper, we assume that the spectra of GRBs are a Band function 
\citep{band1993} in which $\alpha=-1$, $\beta=-2.5$ and
$\eop=\ep/(1+z)$.  We have also done simulations assuming $\alpha =
-0.5$ and -1.5, and $\beta = -2.0$ and -3.0; these different choices
make little difference in our results.

Given $\eiso$, $\ep$, and $T$, we calculate $L_{\rm iso}^{E} = \eiso/T$
and the normalization constant $A$ of the Band spectrum in the rest
frame of the burst source.  We can then calculate the following peak
fluxes and fluences:
\beq
\fe = \frac{L_{\rm iso}^{E}}{4\pi r^{2}(z) (1+z)^{2}} 
	= [\rm erg\; cm^{-2}\; s^{-1}] \; ;
\;\;\;\;\;\;\;\;\;
\se = \fe \cdot T \cdot (1+z) = [\rm erg\; cm^{-2}] \; ;
\eeq
\beq
\fn = \frac{L_{\rm iso}^{N}}{4\pi r^{2}(z) (1+z)} 
	= [\rm phot\; cm^{-2}\; s^{-1}] \; ;
\;\;\;\;\;\;\;\;\;
\sn = \fn \cdot T \cdot (1+z) = [\rm phot\; cm^{-2}] \; .
\eeq
However, these are bolometric quantities, not observed quantities; in
order to calculate the observed peak fluxes and fluences, we must model
the instruments.

Given $\eiso$, $\ep$, and $z$ from the simulations, we calculate the
normalization constant $A^{*}$ of the Band function by considering the
bolometric fluence as observed in our reference frame.
\beq
\se^{\rm bol} = \frac{\eiso}{4\pi (1+z) r(z)^{2}} = A^{*} \times 
	\int_{0}^{\infty}\, EN(E; \alpha,\beta,\eop,A=1)\,dE
\eeq

Once we have $A^{*}$, we can calculate the observed fluxes and fluences
in the passband of our instrument,
\beq
S_{E}^{\rm obs} = \int_{\rm instr}\, EN(E; \alpha,\beta,\eop,A^{*})\,dE \; ;
\;\;\;\;\;\;\;\;\;
F_{E}^{P,{\rm obs}} = \frac{S_{E}^{obs}}{T(1+z)} \; ;
\label{SE_obs}
\eeq
\beq
S_{N}^{\rm obs} = \int_{instr}\, N(E; \alpha,\beta,\eop,A^{*})\,dE \; ;
\;\;\;\;\;\;\;\;\;
F_{N}^{P,{\rm obs}} = \frac{S_{N}^{obs}}{T(1+z)} \; .
\label{SN_obs}
\eeq

To determine whether a particular burst will be detected by a particular
instrument, we define the efficiency as a function of $\eop$,
\beq
\epsilon (\eop) = 
	\frac{\int_{\rm instr}\, N(E; \alpha,\beta,\eop,A^{*})\,dE}
	{\int_{0.1}^{10000}\, N(E; \alpha,\beta,\eop,A^{*})\,dE}
	= \frac{F_{N}^{P,{\rm inst}}}{F_{N}^{P,\oplus}} \; ,
\eeq
where $F_{N}^{P,\oplus}$ and $F_{N}^{P,{\rm inst}}$ are the bolometric
peak photon number flux of the burst at the Earth and the peak photon
number flux of the burst as measured by a particular instrument,
respectively.  This expression gives a shape function which we
normalize to Figures 2 through 9 of \cite{band2003} for the desired
detector.  Note that our shape function is the same as Band's, except
that we consider incident burst spectra extending from 0.1 - 10,000 keV
instead of from 1 - 1000 keV, in order to encompass the full range of
values of $\eop$ observed by HETE-2 for XRFs, X-ray-rich GRBs, and GRBs.

The normalization is approximately given by
\beq
D_{inst}=\frac{C_{\rm min}(\sigma,B)}{A_{\rm eff} \cdot \Delta t_{\rm trig}}
	= [\rm phot\; s^{-1}\; cm^{-2}] \; ,
\eeq
where $C_{\rm min}(\sigma,B)$ is the minimum detectable number of counts in
the detector, $\sigma$ is the SNR required for detection, $B$ is the
background count rate from the diffuse X-ray background, $A_{\rm eff}$ is the
effective area of the detector, and $\Delta t_{\rm trig}$ is the trigger
timescale \citep{band2003}.  A burst is detected if
\beq
F_{N}^{P,\oplus} \ge \frac{D_{\rm inst}}{\epsilon (\eop)} 
	= F_{N,{\rm min}}^{P,\oplus} \; ;{\rm \ i.e.,}
\;\;
F_{N}^{P,{\rm inst}} \ge D_{\rm inst} = F_{N,{\rm min}}^{P,{\rm inst}}
\; .
\eeq
Thus $D_{\rm inst}$ is the peak number flux detection threshold in the
instrument passband.

We have reproduced the results of \cite{band2003} for BATSE on CGRO,
the WFC and GRBM on \bsax, and the WXM and FREGATE on HETE-2.  However,
we use a trigger timescale $\Delta t_{\rm trig} = 5$ seconds for the
WXM on HETE-2, rather than the value of 1 second used by
\cite{band2003}.  We also use a threshold SNR for detection of a burst
by the GRBM on \bsax of 15 \citep{costa2003}, rather than the value of
$5.6$ used by \cite{band2003}.\footnote{The reason for this is the
following:  The half opening angle of the WFC is $\theta_{\rm WFC} =
20^\circ$.  The GRBM consists of four anti-coincidence shields, two of
which are normal to the WFC boresight and two of which are parallel to
it.  In order to be detected, a burst must be detected in at least two
of the anti-coincidence shields; i.e., it must exceed 5 $\sigma$ in one
of the anti-coincidence shield that is normal to the WFC boresight {\it
and} in one of the two anti-coincidence shields that are parallel to
the WFC boresight.  A burst that exceeds 5 $\sigma$ in one of the two
anti-coincidence shields that are parallel to the WFC boresight and is
localized by the WFC (i.e., that lies within 20$^\circ$ of the WFC
boresight) exceeds 25 $\sigma$ in the anti-coincidence shield that is
normal to the WFC boresight.  Detailed Monte Carlo simulations show
that some of a burst's gamma-rays are scattered by material in the WFC
into one or the  other of the two anti-coincidence shields that are
parallel to the WFC boresight.  This reduces the required SNR of the
burst in the anti-coincidence shield that is normal to the WFC
boresight to $\approx 15 \sigma$ \citep{costa2003}.}

Figure \ref{uniform_jet_hete_bsax} shows the threshold sensitivity
curves in peak photon number flux $\fn$ for the WXM and FREGATE on
HETE-2 and for the WFC and GRBM on \bsax as a function of $\eop$, the
observed peak energy of the $\nu F_\nu$ spectrum of the burst.  Since
\bsax could not trigger on WFC data and was forced to rely on the
less-sensitive GRBM for its triggers, we consider a burst to have been
detected by \bsax only if its peak flux falls above the GRBM
sensitivity threshold.  Since \hetetwo can trigger on WXM data, we
consider a burst to have been detected if its peak flux falls above the
minimum of the WXM and FREGATE sensitivity thresholds.  These bursts
form the ensemble of observed bursts from which we construct various
observed distributions.

\section{Results}

In comparing the observed properties of XRFs, X-ray-rich GRBs, and
GRBs, and their predicted properties in the variable jet opening-angle
model, we consider values of the power-law index for the distribution
of jet solid angles $\ojet$ of $\delta$ = 1, 2, and 3.  As we will see,
the observed properties of the bursts are fit best by $\delta = 2$,
which implies approximately equal numbers of bursts per logarithmic
interval in all  observed quantities.

In comparing the observed properties of XRFs, X-ray-rich GRBs, and
GRBs, and their predicted properties in the power-law universal jet
model, we adopt $\eiso \propto \oview^{-1}$ since this relation is
required in order to recover the standard energy result for GRBs.  In
addition, we consider two possibilities for the range of $\oview$.  In
the first case, we require the power-law universal jet model to account
for the full range of the $\eiso-\ep$ relation, including XRFs,
X-ray-rich GRBs, and GRBs; i.e., we fix the normalization of
$\egamma^{0}$ so that the smallest value of $\eiso$ given by the model
is the value of $\eiso$ for XRF 020903.  In the second case, we fix the
normalization of $\egamma ^{0}$ so that the smallest  value of $\eiso$
given by the model is the $\eiso$ value for GRB 980326, the GRB with
the smallest $\eiso$ in the \bsax sample.

The data sets for $\se$ and $\eop$, and especially for $\eiso$ and
$\ep$ (which require knowledge of the redshift of the burst), are
sparse at the present time.  The latter two data sets also suffer from
a large observational selection effect (there is a dearth of XRFs with
known redshifts because the X-ray and optical afterglows of XRFs are so
faint).  In addition, the KS test (which is the appropriate test to use
to compare cumulative distributions) is notoriously weak.  We therefore
do not think that it is justified to carry out detailed fits to these
data sets at this time -- in fact we think that doing so is likely to
produce highly misleading results.  We have therefore contented
ourselves with making fits to these data sets ``by eye,'' which can
support qualitative -- but not quantitative -- conclusions.


Figure \ref{uniform_jet_hete_bsax} shows the detectability of bursts by
HETE-2 and \bsax in the variable jet opening-angle model for $\delta =
2$.  Detected bursts are shown in blue and non-detected bursts in red. 
The left-hand panels show bursts in the [$\eiso,\ep$]-plane detected by
HETE-2 (upper panel) and by \bsax (lower panel).  For each experiment,
we overplot the locations of the HETE-2 and \bsax bursts with known
redshifts.  The observed burst in the lower left-hand corner of the
HETE-2 panel is XRF 020903, the most extreme burst in our sample.  The
agreement between the observed and predicted distributions of bursts is
good.  The right-hand panels show bursts in the [$\eop, \fn(0.1-10000\
{\rm keV})$]-plane detected by HETE-2 (upper panel) and by \bsax (lower
panel).  For each experiment we show the sensitivity thresholds for
their respective instruments plotted in solid blue.  The BATSE
threshold is shown in both panels as a dashed blue line.  Again, the
agreement between the observed and predicted distributions of bursts is
good.  The left-hand panels exhibit the constant density of bursts per
logarithmic interval in $\eiso$ and $\ep$ given by the variable jet
opening-angle model for $\delta = 2$.  Since $\liso =  \eiso/T$, this
choice of $\delta = 2$ corresponds to a GRB luminosity function
$f(\liso) \propto \liso^{-1}$, which is roughly consistent with those
found by \cite{schmidt2001} and \cite{lloyd-ronning2002}.


Figure \ref{SE_SN_vs_omega} shows scatter plots of $\se$ and $\sn$
versus $\ojet$.  The top panels show the predicted distributions in the
variable jet model for $\delta = 2$, while the bottom panels show the
power-law universal jet opening-angle model pinned to the $\eiso$
value of XRF 020903.  Detected bursts are shown in blue and
non-detected bursts in red.  The top panels exhibit the constant
density of bursts per logarithmic interval in $\se$, $\sn$, and $\ojet$
given by the variable jet opening-angle model for $\delta = 2$.  The
bottom panels exhibit the concentration of bursts at $\ojet \equiv
\oview \approx 2 \pi$ and the resulting preponderance of XRFs relative
to GRBs in the power-law universal jet model when it is pinned to
the $\eiso$ value of XRF 020903; i.e., when one attempts to extend the
model to include XRFs and X-ray-rich GRBs, as well as GRBs.


Figure \ref{FEP_FNP_vs_omega} shows scatter plots of $\fe$ and $\fn$
versus $\ojet$.  The top panels show the predicted distributions in the
variable jet opening-angle model for $\delta = 2$, while the middle and
the bottom panels show the power-law universal jet model pinned to
the $\eiso$ values of XRF 020903 and GRB 980326, respectively.   In
these scatter plots, as in the other scatter plots presented in this
paper, we show a random subsample (usually 5000 bursts) of the 50,000
bursts that we have generated.  Detected bursts are shown in blue and
non-detected bursts in red.  The top panels exhibit the constant
density of bursts per logarithmic interval in $\fe$, $\fn$, and $\ojet$
given by the variable jet opening-angle model for $\delta = 2$.  The
middle and bottom panels show the concentration of bursts at $\ojet
\equiv \oview \approx 2 \pi$.  The middle panels show the resulting
preponderance of XRFs relative to GRBs in the power-law universal jet
model when it is pinned to the $\eiso$ values of XRF 020903; i.e.,
when one attempts to extend the model to include XRFs and X-ray-rich
GRBs, as well as GRBs.


Figure \ref{omega_cuml} shows the observed and predicted cumulative
distributions of $\ojet$.  The left panel shows the cumulative
distributions of $\ojet$ predicted by the variable jet opening-angle
model for $\delta = 1$, $2$ and $3$ (solid curves), compared to the
observed cumulative distribution of the values of $\ojet$ given in
\cite{bloom2003} scaled downward by a factor of $C_{\rm jet}$ = 95
(solid histogram).  The predicted cumulative distribution of $\ojet$
given by $\delta = 2$ fits the the shape and values of the scaled
$\ojet$ distribution reasonably well.  The right panel shows the
cumulative $\ojet \equiv \oview$ distributions predicted by the
power-law universal jet model with the minimum value of $\eiso$
pinned to the value of $\eiso$ for XRF 020903 (solid curve) and to
the value of $\eiso$ for GRB 980326 (dashed curve)  These models are
compared with the observed cumulative distribution of the values of
$\ojet$ given in \cite{bloom2003} (dashed histogram) and the same
distribution scaled downward by a factor of $C_{\rm jet}$ = 95 (solid
histogram).  The cumulative $\ojet$ distribution predicted by the
power-law universal jet model pinned to GRB 9980326 fits the shape
and values of the observed cumulative distribution given by the values
of $\ojet$ in \cite{bloom2003} reasonably well if the observed values
are scaled upward by a factor of $\approx$ 7.  The cumulative
distribution of $\ojet$ predicted by the power-law universal jet model
pinned to XRF 020903 does not fit the shape of the observed
cumulative distribution of $\ojet$ for any scaling factor.


Figure \ref{3models_scatter} shows scatter plots of $\eiso$ versus
$\ep$ (left column) and  $\ep$ versus $\se$ (right column).  The top
panels show the predicted distributions in the variable jet
opening-angle model for $\delta = 2$, while the middle and the bottom
panels show the power-law universal jet model pinned to the $\eiso$
values of XRF 020903 and GRB 980326, respectively.  Detected bursts are
shown in blue and non-detected bursts in red.  In the left column, the
black triangles and circles show the locations of the \bsax and HETE-2
bursts with known redshifts.  In the right column, the black cirlces
show the locations of HETE-2 bursts for which joint fits to WXM and
FREGATE spectral data have been carried out \citep{sakamoto2003b}.  The
top panels exhibit the constant density of bursts per logarithmic
interval in $\eiso$, $\ep$, and $\se$ given by the variable jet
opening-angle model for $\delta = 2$.  The middle and bottom panels
show the limited range in $\eiso$, $\ep$, and $\se$ of detected bursts
in the power-law universal jet model.  The middle panels show the
preponderance of XRFs relative to GRBs predicted in the power-law
universal jet model when it is pinned to the $\eiso$ value of XRF
020903; i.e., when one attempts to extend the model to include XRFs and
X-ray-rich GRBs, as well as GRBs.


Figure \ref{fab_four} compares the observed and predicted cumulative
distributions of $\eiso$ and $\ep$ for \bsax and HETE-2 bursts with
known redshifts, and the observed and predicted cumulative
distributions of $\se$ and $\eop$ for all HETE-2 bursts.  The solid
histograms are the observed cumulative distributions.  The solid curves
are the cumulative distributions predicted by the variable jet
opening-angle model for $\delta = 2$.  The dotted curves are the
cumulative distributions predicted by the power-law universal jet model
pinned at the $\eiso$ value of XRF 020903; i.e., when one attempts
to extend the model to include XRFs and X-ray-rich GRBs, as well as
GRBs.  The dashed curves are the cumulative distributions predicted by
the power-law universal jet model pinned at the $\eiso$ value of
GRB 980326; i.e., when one fits the model only to GRBs.  The cumulative
distributions in the present figure correspond to those formed by
projecting the observed and predicted distributions in Figure
\ref{3models_scatter} onto the $x$- and $y$-axes of the panels in that
figure.  The present figure shows that variable jet opening-angle model
for $\delta = 2$ can explain the observed distributions of burst
properties reasonably well, especially given that the sample of XRFs
with known redshifts is incomplete due to optical observational
selection effects (see Section 6.6.1).  It also shows that the
power-law universal jet model can explain the observed distributions of
GRB properties reasonably well, but cannot do so if asked to explain
the properties of XRFs and X-ray-rich GRBs, as well as GRBs.


Figure \ref{FNP_EPO_z1_scatter} shows scatter plots of $\fn$ (left
column) and $\eop$ (right column) as a function of redshift.  The top
row shows the distributions of bursts predicted by the variable jet
opening-angle model for $\delta = 2$.  The middle row shows the
distributions of bursts predicted by the power-law universal jet model
pinned to the $\eiso$ value for XRF 020903, while the bottom row
shows the distributions of bursts predicted by the power-law universal
jet model pinned to the $\eiso$ value for GRB 980326.  Detected
bursts are shown in blue and non-detected bursts in red.  The black
circles show the positions of the HETE-2 bursts with known redshifts. 
This figure shows that variable jet opening-angle model for $\delta =
2$ can explain the observed distributions of bursts in the ($1+z,\fn$)-
and ($1+z,\eop$)-planes reasonably well.  It also shows that the
power-law universal jet model can explain the observed distributions of
GRBs alone reasonably well, but cannot explain the observed
distributions of XRFs, X-ray-rich GRBs, and GRBs.  These conclusions
are confirmed by Table \ref{kind_table}, which shows the percentages of
XRFs, X-ray-rich GRBs, and hard GRBs in the HETE-2 data, and predicted
by the variable jet opening-angle model and the power-law universal jet
model.


Figure \ref{Lgamma_scatter_cuml} shows scatter plots of $\liso$ versus
$\ep$ and a comparison of the observed and predicted cumulative
distributions of $\liso$.   The upper left panel shows the distribution
of bursts predicted by the variable jet opening-angle model for $\delta
= 2$.  The lower left panel shows the distribution of bursts predicted
by the power-law universal jet model pinned at the $\eiso$ value
for XRF 020903.  The lower right panel shows the power-law universal
jet model pinned at the $\eiso$ value for GRB 980326.  Detected
bursts are shown in blue and non-detected bursts in red.  The black
circles show the positions of the HETE-2 bursts with known redshifts. 
The upper right panel shows the observed cumulative distribution of
$\liso$ for HETE-2 bursts with known redshifts (histogram) compared
with the cumulative $\liso$ distribution predicted by the variable jet
opening-angle model for $\delta = 2$ (solid curve), and the cumulative
$\liso$ distributions predicted by the power-law universal jet model
pinned at the $\eiso$ value for XRF 020903 (dotted curve) and for
GRB 980326 (dashed curve).  The figure shows that the variable jet
opening-angle model for $\delta = 2$ can explain the observed
cumulative distributions of bursts in the ($\liso,\ep$)-plane
reasonably well.  It also shows that the power-law universal jet model
can explain the observed distribution of $\liso$ for GRBs alone
reasonably well, but cannot explain the observed distribution for XRFs,
X-ray-rich GRBs, and GRBs.


The left panel of Figure \ref{FEP_dists} shows the observed cumulative
distribution of $\fe$ for HETE-2 bursts (histogram) compared with the
cumulative $\fe$ distribution predicted by the variable jet
opening-angle model for $\delta = 2$ (solid curve), and the cumulative
$\fe$ distributions predicted by the power-law universal jet model
pinned at the $\eiso$ value for XRF 020903 (dotted curve) and for
GRB 980326 (dashed curve).  This figure shows that variable jet
opening-angle model for $\delta = 2$ can explain the observed
cumulative distribution of $\fe$ for HETE-2 bursts reasonably well.  It
also shows that the power-law universal jet model can explain the
observed cumulative distribution of $\fe$ for GRBs alone reasonably
well, but cannot explain the observed distribution for XRFs, X-ray-rich
GRBs, and GRBs seen by HETE-2.  All three models have some difficulty
explaining the cumulative $\fe$  distribution for BATSE bursts
\citep{donaghy2003b}.  The right panel of Figure \ref{FEP_dists} shows
the differential distribution of $\eop$ predicted by the variable jet
opening-angle model for $\delta = 2$ for bursts with $\fe > 10^{-8}$
(solid histogram), $10^{-7}$ (dashed histogram), and $ 10^{-6}$ erg
cm${-2}$ s$^{-1}$ (dotted histogram).  The last distribution is in rough
agreement with that found by \cite{preece2000} for BATSE bursts with
$\fe \gtrsim 5 \times 10^{-7}$ erg cm$^{-2}$ s$^{-1}$ and $\se > 4
\times 10^{-5}$ erg cm$^{-2}$.

\section{Discussion}

\subsection{Structure of GRB Jets}

Motivated by the HETE-2 results, we have explored in this paper the
possibility of a unified jet model of XRFs, X-ray-rich GRBs, and GRBs. 
The HETE-2 results show that $\se$ and $\eiso$ decrease by a factor
$\sim 10^5$ in going from GRBs to XRFs (see Figures
\ref{hete_Epobs_flu_by_class} and \ref{amati_plots}).  Figures 13 - 16
show that the variable jet opening-angle model can accomodate the large
observed ranges in $\se$ and $\eiso$ reasonably well, while the
power-law universal jet model cannot.

Figures 13-16 show that the variable jet opening-angle model with
$\delta = 2$ can explain a number of the observed properties of GRBs
reasonably well.  These figures show that the power-law universal jet
model \citep{rossi2002,zhang2002,meszaros2002,wzhang2004,perna2003}
with $\eiso \propto \thetav^{-2} \propto \oview^{-1}$
\citep{zhang2002,rossi2002,perna2003} can also explain a number of the
observed properties of GRBs reasonably well [see also
\cite{rossi2002,perna2003}].

However, as we have seen, HETE-2 has provided strong evidence that the
properties of XRFs, X-ray-rich GRBs, and GRBs form a continuum in the
($\eiso,\ep$)-plane \citep{lamb2003} and in the ($\se,\eop$)-plane
\citep{sakamoto2003b}, and therefore that these three kinds of bursts
are the same phenomenon.  If this is true, it implies that the 
$\egamma$ inferred by \cite{frail2001}, \cite{panaitescu2001} and
\cite{bloom2003} is too large by a factor of at least 100.  The reason
is that the values of $E_{\rm iso}$ for XRF 020903
\citep{sakamoto2003a} and XRF 030723 \citep{lamb2003} are $\sim$ 100
times smaller than the value of $\egamma$ inferred by \cite{frail2001}
and \cite{panaitescu2001} -- an impossibility.

The reason is that the predictions of the variable jet opening-angle
and power-law universal jet models differ dramatically if they are
required to accomodate the large observed ranges in $\se$ and
$\eiso$.   Taking $N(\Omega_{\rm jet}) \sim \Omega_{\rm jet}^{-2}$
(i.e., $\delta = 2$), the variable jet opening-angle model predicts
equal numbers of bursts per logarithmic decade in $\se$ and $\eiso$,
which is exactly what HETE-2 sees \citep{lamb2003,sakamoto2003b}(see
Figures \ref{3models_scatter} and \ref{fab_four}).  On the other hand,
in the power-law universal jet model the probability of viewing the jet
at a viewing angle $\thetav$ is $d\oview$, where $\oview$ is the solid
angle contained within the angular radius $\thetav$.  Consequently,
most viewing angles $\thetav$ will be $\theta_{\rm view, max}$ or
$\approx 90^\circ$, whichever is smaller.  This implies that the number
of XRFs should exceed the number of GRBs by many orders of magnitude,
something that HETE-2 does not observe (again, see Figures
\ref{3models_scatter} and \ref{fab_four}).  

Threshold effects can offset this prediction of the power-law universal
jet model over a limited range in $\se$ and $\eiso$.  This is what
enables the power-law universal jet model to explain a number of the
observed properties of GRBs reasonably well
\citep{rossi2002,perna2003}.  However,  threshold effects cannot offset
this prediction over a large range in $\se$ and $\eiso$, as our
simulations confirm.  This is why the power-law universal jet model
cannot accomodate the large observed ranges in $\se$ and $\eiso$.

We conclude that, if $\se$ and $E_{\rm iso}$ span ranges of $\sim
10^5$, as the HETE-2 results strongly suggest, the variable jet
opening-angle model can provide a unified picture of XRFs and GRBs,
whereas the power-law universal jet model cannot.  Thus XRFs may
provide a powerful probe of GRB jet structure.

\subsection{Rate of GRBs and the Nature of Type Ic Supernovae}

A range in $\eiso$ of $10^5$, which is what the HETE-2 results strongly
suggest, requires a {\it minimum} range in $\Delta \ojet$ of $10^4 -
10^5$ in the variable jet opening-angle model.  Thus the unified
picture of XRFs and GRBs in the variable jet opening-angle model
implies that the total number of bursts is
\beq
N_{\rm total} = - \int^{\ojetmax}_{\ojetmin} d\ojet \ojet^{-2} 
\approx (\ojetmin)^{-1} \; .
\label{grb_number}
\eeq
Thus there are $2 \pi/\ojetmin \sim 10^5$ more bursts with very small
$\ojet$'s for every burst that is observable; i.e., the rate of GRBs
may be $\sim 100$ times greater than has been thought.

In addition, since the observed ratio of the rate of Type Ic SNe to the
rate of GRBs in the observable universe is $R_{\rm Type\ Ic}/ R_{\rm
GRB} \sim 10^5$ \citep{lamb1999,lamb2000}, the variable jet
opening-angle model implies that the rate of GRBs could be comparable
to the rate of Type Ic SNe.  More spherically symmetric jets yield XRFs
and narrower jets produce GRBs.  Thus low $E_{\rm peak}$ (intrinsically
faint) XRFs may probe core collapse supernovae that produce wide jets,
while high $E_{\rm peak}$ (intrinsically luminous) GRBs may probe core
collapse supernovae that produce very narrow jets (possibly implying
that the cores of the progenitor stars of these bursts are rapidly
rotating).  

Thus XRFs and GRBs may provide a combination of GRB/SN samples that
would enable astronomers to study the relationship between the degree
of jet-like behavior of the GRB and the properties of the supernova
(brightness, polarization $\Leftrightarrow$ asphericity of the
explosion, velocity of the explosion $\Leftrightarrow$ kinetic energy
of the explosion, etc.).  GRBs may therefore provide a unique
laboratory for understanding Type Ic core collapse supernovae.

\subsection{Constraints on $\ojetmin$ and $\ojetmax$}

The HETE-2 results require a range in $\ojet$ of $\sim 10^5$ within the
context of the variable jet opening-angle model in order to explain the
observed  ranges in $\se$ and $\eiso$.  Thus the HETE-2 results fix the
ratio $\ojetmax/\ojetmin$, but not $\ojetmin$ and $\ojetmax$
separately. However, geometry and observations strongly constrain the
possible values of $\ojetmin$ and $\ojetmax$.  In this paper, we have
adopted $\ojetmax = 0.6 \times 2\pi\ {\rm sr}$ (i.e., $\thetajet =
70^\circ$), which is nearly  the maximum value allowed by geometry. 
However, it seems physically unlikely that GRB jets can have jet
opening angles as large as $\approx 90^\circ$.  One might therefore
wish to adopt a smaller value of $\ojetmax$.  This would imply a
smaller value of $\ojetmin$ and therefore a larger GRB rate.  But the
GRB rate cannot be larger than the rate of Type Ic SNe.  Therefore
$\ojetmin$ cannot be much smaller than the value $\ojetmin = 0.6 \times
2 \pi \times 10^{-5}\ {\rm sr} = 3.8 \times 10^{-5}\ {\rm sr}$ that we
have adopted.

Even the value $\ojetmin = 3 \times 10^{-5}\ {\rm sr}$ implies GRB jet
opening solid angles that are a factor of $\approx$ 100 smaller than
those inferred from jet break times by \cite{frail2001},
\cite{panaitescu2001} and \cite{bloom2003}.  There is a substantial
uncertainty in the jet opening solid angle implied by a given jet break
time, but the uncertainty is thought to be a factor of $\sim 20$, not a
factor $\sim 100$ \citep{rhoads1999,sari1999}.  In addition, the global
modeling of GRB afterglows is largely free from this uncertainty.  Such
modeling tends to find jet opening angles $\thetajet$ of a few degrees
for the brightest and hardest GRBs
\citep{panaitescu2001,panaitescu2003} -- values of $\thetajet$ that are
a factor of at least 3, and in some cases a factor of 10, larger than
the jet opening angles we use in this work.  This is discomforting;
adopting a still smaller value of $\ojetmin$ would be even more
discomforting.

Another constraint on $\ojetmin$ comes from the monitoring of the
late-time radio emission of a sample of 33 nearby Type Ic SNe that has
been carried out by \cite{berger2003c}.  They find that the energy
emitted at radio wavelengths by this sample of Type Ic SNe is $\eradio
< 10^{48}$ ergs in almost all cases.  This implies that these
supernovae do not produce jets with energies $\ejet > 10^{48}$ ergs,
and therefore that at most $\sim$ 4\% of all nearby Type Ic SNe 
produce GRBs, assuming $\egamma = 1.4 \times 10^{51}$ ergs.  In the
variable jet opening-angle model, $\egamma$ is a factor $\approx$ 100
times less than this value, which weakens the constraint on the allowed
fraction of Type Ic SNe that produce GRBs to $\lesssim$ 10\%.  Adopting
a still smaller value of $\ojetmin$ would decrease $\egamma$ and
therefore increase the allowed fraction of Type Ic SNe that produce
GRBs.  However, a smaller value of $\ojetmin$ would also increase the
predicted numbers (and therefore the fraction) of Type Ic SNe that
produce GRBs.  Thus, while not yet contradicting the variable jet
opening-angle model of XRFs and GRBs, the radio monitoring of nearby
Type Ic SNe carried out by \cite{berger2003c} places an important
constraint on $\ojetmin$.

In Section 6.6.1, we report tantalizing evidence that the
efficiency with which the kinetic energy in the jet is converted into
prompt emission at X-ray and $\gamma$-ray wavelengths may decrease as
$\eiso$ and $\ep$ decreases; i.e., this efficiency may be less for XRFs
than for GRBs [see also \cite{lloyd-ronning2004}].  Since
$\eiso$ spans five decades in going from XRFs to GRBs, even a modest
rate of decline in this efficiency with $\eiso$ would reduce the
required range in $\Delta \ojet$ by a factor of ten or more.  Such a
factor would allow $\ojetmin$ to be increased to $\ojetmin \approx 4
\times 10^{-4}$ sr or more, which would bring the jet opening-angle for
GRBs in the variable jet opening angle model into approximate agreement
with the values derived from global modeling of GRB afterglows.  It
would also reduce the predicted rate of GRBs by a factor of ten or
more, and therefore also reduce the fraction of Type Ic SNe that
produce GRBs to $\sim$ 10\% or less, in agreement with the constraint
derived from radio monitoring of nearby Type Ic SNe.  We note that
including such a decrease of efficiency with $\eiso$ and $\ep$ would
introduce an additional parameter into the model.

\subsection{Outliers}

\cite{bloom2003} and \cite{berger2003b} have called attention to the
fact that not all GRBs have values of $\egamma$ that lie within a
factor of 2-3 of the standard energy $\egamma$; i.e., that there
are outliers in the $\egamma$ distribution.  \cite{berger2003c} have
also proposed a core/halo model for the jet in GRB 030329.

In addition, we note that the two XRFs for which redshifts or strong
redshifts constraints exist (the HETE--localized bursts  XRF 020903 and
XRF 030723) lie squarely on the relation between $\eiso$ and $\ep$
found by \cite{amati2002} (see Figure \ref{amati_plots}).  The implied
value of $\egamma$ from the absence of a jet break in the optical
afterglow of XRF 020903 is $\approx 1.1 \times 10^{49}$ ergs
\citep{soderberg2003}, which is consistent with the standard energy of
$\egamma  = 1.17 \times 10^{49}$ ergs that we use in this work. 
However, the implied value of $\egamma$ from the jet break time of
$\sim$ 1 day in XRF 030723 \citep{dullighan2003} is a factor $\sim 100$
smaller than the standard energy that we use in this work and a factor
$\sim 10^4$ smaller than the standard energy of $\egamma = 1.3 \times
10^{51}$ found by \cite{bloom2003} [see also \cite{frail2001} and
\cite{panaitescu2001}].

The unified jet model of XRFs, X-ray-rich GRBs, and GRBs that we have
proposed is a phenomenological one, and is surely missing important
aspects of the GRB jet phenomenon, which may include a significant
stochastic element.  It therefore cannot be expected to account for the
properties of all bursts.  Only further observations can say whether
the bursts discussed above (or others) are a signal that the unified
jet model is missing important aspects of GRB jets, or whether they are
truly outliers.

\subsection{Variable Jet Opening-Angle Model in the MHD Jet Picture}

\cite{zhang2003} and \cite{kumar2003} have studied the early
afterglows of two GRBs.  \cite{zhang2003} find in the case of GRB
990123 strong evidence that the jet is magnetic energy dominated;
\cite{kumar2003} reach a similar conclusion for GRB 021211.  In both
cases, it appears that the magnetic energy dominated the kinetic
energy in the ejected matter by a factor $>$ 1000.  The recent
discovery that the prompt emission from GRB 021206 was strongly
polarized \citep{coburn2003} may provide further support for the
picture that GRBs come from magnetic-energy dominated jets.

Part of the motivation for the power-law universal jet model comes from
the expectation that in hydrodynamic jets, entrainment and the
interaction of the ultra-relativistic outflow with the core of the
progenitor star may well result in a strong fall-off of the velocity of
the flow away from the jet axis.  Thus the narrow jets we find in the
unified picture  of XRFs, X-ray-rich GRBs and GRBs based on the
variable jet opening-angle model are difficult to reconcile with
hydrodynamic jets.  They may be much easier to understand if GRB jets
are magnetic-energy dominated; i.e., if GRBs come from MHD jets.  Such
jets can be quite narrow \citep{vlahakis2001,proga2003,fendt2003} and
may resist the entrainment of material from the core of the progenitor
star.

\subsection{Possible Tests of the Variable Jet Opening-Angle Model}

\subsubsection{X-Ray and Gamma-Ray Observations}

We have shown that a unified picture of XRFs and GRBs based on the
variable jet opening-angle model has profound implications for the
structure of GRB jets, the rate of GRBs, and the nature of Type Ic
supernovae.  Obtaining the evidence needed to confirm (or possibly 
rule out) the variable jet opening-angle model and its implications
will require the determination of both the spectral parameters and the
redshifts of many more XRFs.  The broad energy range of HETE-2 (2-400
keV) means that it is able to accurately determine the spectral
parameters of the XRFs that it localizes.  This will be more difficult
for {\it Swift}, whose spectral  coverage (15-150 keV) is more
limited. 

Until very recently, only one XRF (XRF 020903; Soderberg et al. 2003)
had a probable optical afterglow and redshift (see Figure
\ref{hete_Epobs_flu_by_z}).  This is because the X-ray (and by
implication the optical) afterglows of XRFs are $\sim 10^3$ times
fainter than those of GRBs (see Figure \ref{LX_Eiso}; see also Lamb et
al. (2005) [in preparation]).  However, we find that the best-fit slope
of the correlation between $\lx$ and $\liso$ is not $+1$, but $+0.74
\pm 0.17$.  This implies that the fraction of the kinetic energy of the
jet that goes into the burst itself decreases as $\liso$ (and therefore
$\eiso$) decreases; i.e., the fraction of the kinetic energy in the jet
that goes into the X-ray and optical afterglow is much larger for XRFs
than it is for GRBs. 

This result is consistent with a picture in which the central engines
of XRFs produce less variability in the outflow of the jet than do the
central engines of GRBs, resulting in less efficient extraction of the
kinetic energy of the jet in the burst itself in the case of XRFs than
in the case of GRBs.  Such a picture is supported by studies that
suggest that the temporal variability of a burst is a good indicator of
the isotropic-equivalent luminosity $\liso$ of the burst
\citep{fenimore2000,reichart2001}.  These studies imply that XRFs,
which are much less luminous than GRBs, should exhibit much less
temporal variability than GRBs.  As we discussed in Section 6.3,
if the efficiency with which the kinetic energy in the jet is converted
into prompt emission at X-ray and $\gamma$-ray wavelengths decreases
even modestly with decreasing $\eiso$, it would reduce the required
range in $\Delta \ojet$ by a factor of ten or more, allowing the
opening angle for GRBs in the variable jet opening-angle model to be
brought into approximate with the values derived from global modeling
of GRB afterglows \citep{panaitescu2001,panaitescu2003} and the rate of
GRBs to be brought into agreement with the constraint derived from
radio monitoring of nearby Type Ic SNe \citep{berger2003c}.

The above picture differs from the core-halo picture of GRB jets
recently proposed by \cite{berger2003b}, in which the prompt burst
emission and the early X-ray and optical afterglows are due to a narrow
jet, while the later optical and the radio afterglows are due to a
broad jet.  In this picture, the total kinetic energy $\ejet$ of the
jet is roughly constant, but the fraction of $\ejet$ that is radiated
in the narrow and the broad components can vary.

The challenge presented by the fact that the X-ray (and by implication
the optical) afterglows of XRFs are $\sim 10^3$ times fainter than
those of GRBs can be met: the recent HETE-2--localization of XRF 030723
represents the first time that an XRF has been localized in real time
\citep{prigozhin2003}; identification of its X-ray (Butler et al.
2003a,b) and optical \citep{fox2003c} afterglows rapidly followed. 
This suggests that {\it Swift}'s ability to rapidly follow up GRBs with
the XRT and UVOT -- its revolutionary feature -- will greatly increase
the fraction of bursts with known redshifts.  

A partnership between HETE-2 and {\it Swift}, in which HETE-2 provides
the spectral parameters for XRFs, and {\it Swift} slews to the
HETE-2--localized XRFs and provides the redshifts, can provide the data
that is required in order to confirm (or possibly rule out) the
variable jet opening-angle model and its implications.  This
constitutes a compelling scientific case for continuing HETE-2 during
the {\it Swift} mission.

\subsubsection{Global Modeling of GRB Afterglows}

\cite{panaitescu2003} have modeled in detail the afterglows of GRBs
990510 and 000301c.  In both cases, they find that fits to the X-ray,
optical, NIR, and radio data for GRBs 990510 and 000301c favor the
variable jet opening-angle model over the power-law universal jet
model.  Detailed modeling of the afterglows of other GRBs may provide
further evidence favoring one phenomenological jet model over the other
for particular bursts.

\subsubsection{Polarization of GRBs and Their Afterglows}

The variable jet opening-angle model and the power-law universal jet
model predict different behaviors for the polarization of the optical
afterglow.  The variable jet opening-angle model predicts that the
polarization angle should change by $180^\circ$ over time, passing
through $0$ around the time of the jet break in the afterglow light
curve.  In contrast, the power-law universal jet model predicts that
the polarization angle should not change with time.  The polarization
data on GRB afterglows that has been obtained  to date is in most cases
very sparse, making it difficult to tell whether or not the behavior of
the polarization favors the variable jet opening-angle or the power-law
universal jet model.

In the case of GRB 021004, however, the data shows clear evidence that
the polarization angle changed by approximately $180^\circ$ and changed
sign at roughly the time of the jet break, as the variable jet
opening-angle model, but not the power-law universal jet model, predicts
\citep{rol2003}.  Thus, in the case of this one GRB, at least, the
behavior of the polarization of the optical afterlow favors the
variable jet opening-angle model over the power-law universal jet
model.

\subsection{Rate of Detection of GRBs by Gravitational Wave Detectors}

If, as the variable jet opening-angle model of XRFs, X-ray-rich GRBs
and GRBs implies, most GRBs are bright and have narrow jets -- possibly
implying that the collapsing core of the progenitor star may be rapidly
rotating -- GRBs might be detectable sources of gravitational waves. 
If as has been argued, $E_{\rm gw}/E_{\rm rot} \sim$ 5\% in the
formation of a black hole from the collapse of the core of the Type Ic
supernova \citep{putten2002}, where $E_{\rm gw}$ is the energy emitted
in gravitational waves and $E_{\rm rot}$ is the rotational energy of
the newly formed black hole, and the rate of GRBs is $\sim$ 100 times
higher than has been thought, then the rate of LIGO/VIRGO detections of
GRBs might be $\sim$ 5 yr$^{-1}$ rather than $\sim$ 1 yr$^{-1}$
\citep{putten2002}.

\section{Conclusions}

In this paper we have shown that a variable jet opening-angle model, in
which the isotropic-equivalent energy $\eiso$ depends on the jet solid
opening angle $\ojet$, can account for many of the observed properties
of XRFs, X-ray-rich GRBs, and GRBs in a unified way.  We have also
shown that, although the power-law universal jet model can account
reasonably well for many of the observed properties of GRBs, it cannot
easily be extended to accommodate XRFs and X-ray-rich GRBs.  The
variable jet opening-angle model implies that the total radiated
energy in gamma rays $\egamma$ is $\sim 100$ times smaller than has
been thought.  The model also implies that the hardest and most
brilliant GRBs have jet solid angles $\ojet/2 \pi \sim 10^{-5}$.  Such
small solid angles are difficult to achieve with hydrodynamic jets, and
lend support to the idea that GRB jets are magnetic-energy dominated. 
Finally, the variable jet opening-angle model implies that there are
$\sim 10^5$ more bursts with very small $\ojet$'s for every observable
burst.  The observed ratio of the rate of Type Ic SNe to the rate of
GRBs is $R_{\rm Type Ic}/ R_{\rm GRB} \sim 10^5$; the variable jet
opening-angle model therefore implies that the GRB rate may be
comparable to the rate of Type Ic SNe, with more spherically symmetric
jets yielding XRFs and narrower jets producing GRBs.  GRBs may
therefore provide a unique laboratory for understanding Type Ic core
collapse supernovae. 

\acknowledgments
This research was supported in part by NASA Contract NAGW-4690 and NASA
Grant NAG5-10759.  We would like to thank John Heise for valuable
conversations about X-Ray Flashes, and Bing Zhang, Peter M\'esz\'aros,
and Elena Rossi for valuable conversations about the nature of GRB
jets.

\clearpage

\begin{deluxetable}{cccc}
\tablecaption{Parameters of Lognormal Distributions}
\tablewidth{0pt}
\tablehead{
\colhead{Quantity} & \colhead{Central Value\tablenotemark{a}}
& \colhead{Sigma\tablenotemark{a}} & \colhead{Source of Data\tablenotemark{b}}
}
\startdata
Energy Radiated: Log $E_{\gamma}$[erg] & $51.070\pm 0.095$ &
$0.33^{+0.08}_{-0.06}$ 
& 1 \\ 
$\eiso$-$\ep$ Relation: Log $C$[kev] & $1.950\pm 0.040$ & $0.13$ 
& 2, 3\\
Conversion Timescale: Log $T$ [sec] & $0.574\pm 0.075$ & $0.305^{+0.062}_{-0.049}$ 
& 2, 3\\
\enddata
\tablenotetext{a}{Lognormal distribution.}
\tablenotetext{b}{References: 1. Bloom et al. (2003); 2. Amati et al. (2002); 3. Lamb et al. (2004d)}
\label{gauss_table}
\end{deluxetable}

\begin{deluxetable}{cccc}
\tablecaption{Percentages of XRFs, X-Ray-Rich GRBs, and
Hard GRBs in HETE-2 Data and Predicted by Different Jet Models}
\tablewidth{0pt}
\tablehead{
\colhead{HETE-2 Data or Model} & \colhead{XRFs (\%)}
& \colhead{X-Ray-Rich GRBs (\%)} & \colhead{Hard GRBs (\%)}
}
\startdata
HETE-2 Data					& $33$ & $44$ & $22$ \\
Variable Opening-Angle Jet			& $22$ & $39$ & $39$ \\
PL Universal Jet pinned to 020903	& $89$ & $10$ & $1.0$ \\
PL Universal Jet pinned to 980326	& $32$ & $56$ & $12$ \\
\enddata
\label{kind_table}
\end{deluxetable}

\clearpage


%
\begin{figure}[ht]  
\begin{center}
\resizebox{5.3cm}{!}{\includegraphics{f2a.eps}}
\resizebox{5.3cm}{!}{\includegraphics{f2b.eps}}
\resizebox{5.3cm}{!}{\includegraphics{f2c.eps}}
\end{center}  
\caption{
{\it Left panel:} distribution of HETE-2 and \bsax bursts in the
($\eiso$,$\liso$)-plane, where $\eiso$ and $\liso$ are the
isotropic-equivalent GRB energy and luminosity in the source frame. 
{\it Middle panel:} distribution of HETE-2 and \bsax bursts in the
($\eiso$,$\ep$)-plane, where $\ep$ is the energy of the peak of the 
burst $\nu F_\nu$ spectrum in the source frame.  The HETE-2 bursts
confirm the relation between $\eiso$ and $\ep$ found by
\cite{amati2002} and extend it by a factor $\sim 300$ in $\eiso$. 
{\it Right panel:} distribution of HETE-2 and \bsax bursts in the
($\liso$,$\ep$)-plane.  
The distribution of HETE-2 and \bsax bursts in the three planes
demonstrates that there is a linear relation between $\log\eiso$ and
$\log\liso$ that extends over at least five decades in both
quantities, and are linear relations between both $\log\eiso$ and
$\log\liso$ and $\log\ep$ that extend over at least 2.5 decades in
$\ep$. The bursts with the lowest and second-lowest values of $\eiso$ 
and $\liso$ are XRFs 020903 and 030723.  From \cite{lamb2003}; \bsax
data is from \cite{amati2002}.
}
\label{amati_plots}
\end{figure}

%
\begin{figure}[t]
\includegraphics[width=6.75truein,clip=]{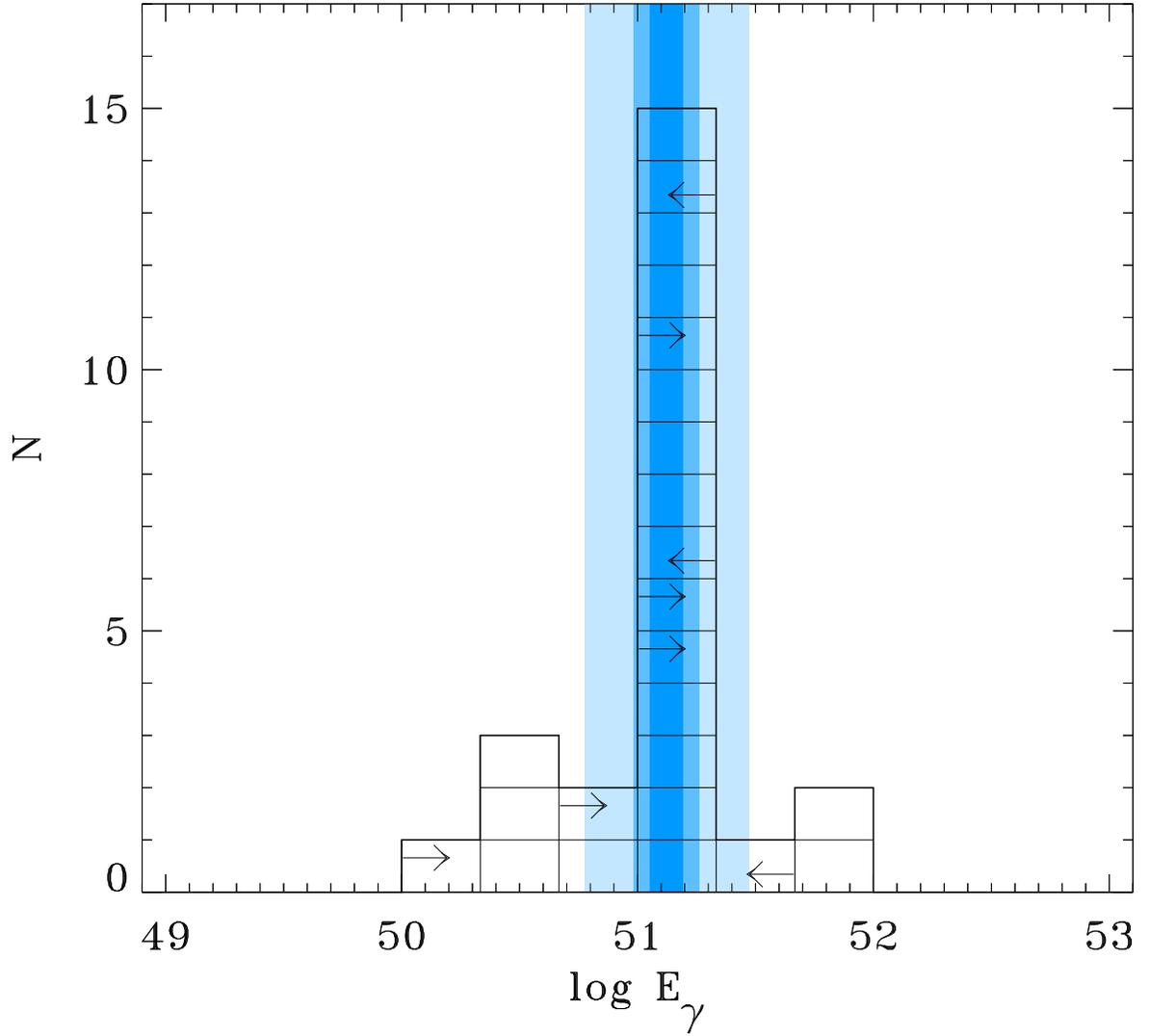}
\caption{Distribution of the total energy $\egamma$ radiated in 
gamma-rays by GRBs, taking into account the jet opening angle
inferred from the jet break time.  From \cite{bloom2003}.
}
\label{bloom_fig1}
\end{figure}

%
\begin{figure}[ht]  
\begin{center}
\resizebox{5.3cm}{!}{\includegraphics{f4a.eps}}
\resizebox{5.3cm}{!}{\includegraphics{f4b.eps}}
\resizebox{5.3cm}{!}{\includegraphics{f4c.eps}}
\end{center}
\caption{
{\it Left panel:} distribution of HETE-2 and \bsax bursts in the ($2
\pi/\ojet$,$\eiso$)-plane, where $\eiso$ is the isotropic-equivalent
burst energy in the source frame.  In the variable jet opening-angle
model, $\ojet$ is the jet solid angle; in the power-law universal jet
model it is the solid angle interior to the viewing angle $\thetav$.
{\it Middle panel:} distribution of HETE-2 and \bsax bursts in the ($2
\pi/\ojet$,$\liso$)-plane, where $\liso$ is the 
isotropic-equivalent burst luminosity in the source frame. 
{\it Right panel:} distribution of HETE-2 and \bsax bursts in the ($2
\pi/\ojet$,$\ep$)-plane, where $\ep$ is the energy of the peak of the
burst $\nu F_\nu$ spectrum in the source frame.  
The distribution of HETE-2 and \bsax bursts in these three planes
demonstrates that there are linear relations between both $\log\eiso$
and $\log\liso$ and $\log\ojet^{-1}$ that extends over at least 2.5
decades in $\log\eiso$ and $\log\liso$, and a relation of slope 1/2
between $\log\ep$ and $\log\ojet^{-1}$ that extends over at least a
decade in $\ep$.  The event in the lower-left corner is XRF 020903,
which is shown as a lower-limit in $2\pi/\ojet$.  Data on $\ojet$ is
from \cite{bloom2003}.
}
\label{ojet_corrs}
\end{figure}

%
\begin{figure}[t]
\includegraphics[width=6.0truein,clip=]{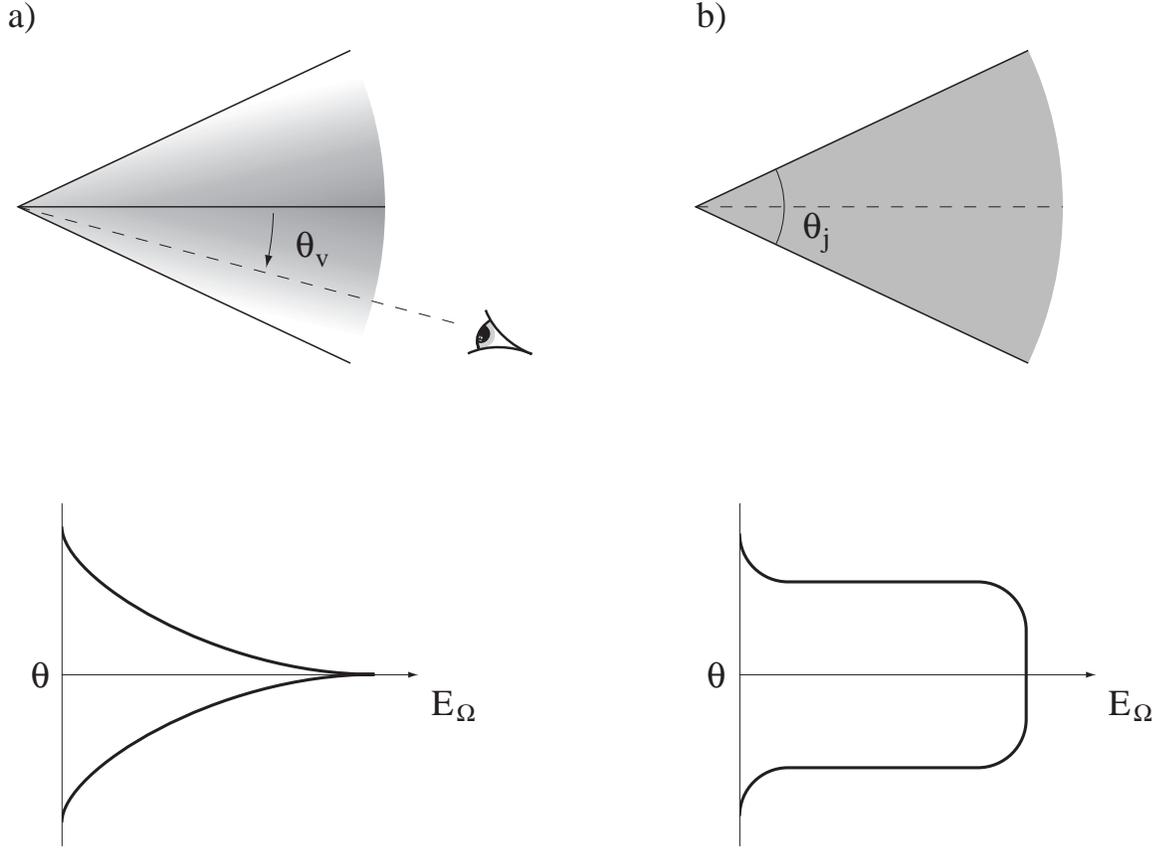} 
\caption{
Schematic diagrams of the power-law universal and variable
opening-angle jet models of GRBs from \cite{ramirez-ruiz2002}.  In the
power-law universal jet model, the isotropic-equivalent energy and
luminosity is assumed to decrease as the viewing angle $\thetav$ as
measured from the jet axis increases.  In order to recover the
standard energy result \citep{frail2001}, $E_{\rm iso} (\thetav)
\sim \thetav^{-2}$ is required.  In the variable jet opening-angle
model, GRBs produce jets with a large range of jet opening angles 
$\theta_{\rm jet}$.  For $\thetav < \theta_{\rm jet}$, $E_{\rm iso}
(\thetav) \approx$ constant while for $\thetav > \theta_{\rm jet}$,
$E_{\rm iso} (\thetav) = 0$.  In this paper, we take $\theta_{\rm jet}$
to be the half-opening angle of the jet.
}
\vskip -0.1truein
\label{two_jet_schematics}
\end{figure}

%
\begin{figure}[ht]  
\begin{center}
\resizebox{!}{20.5cm}{\includegraphics{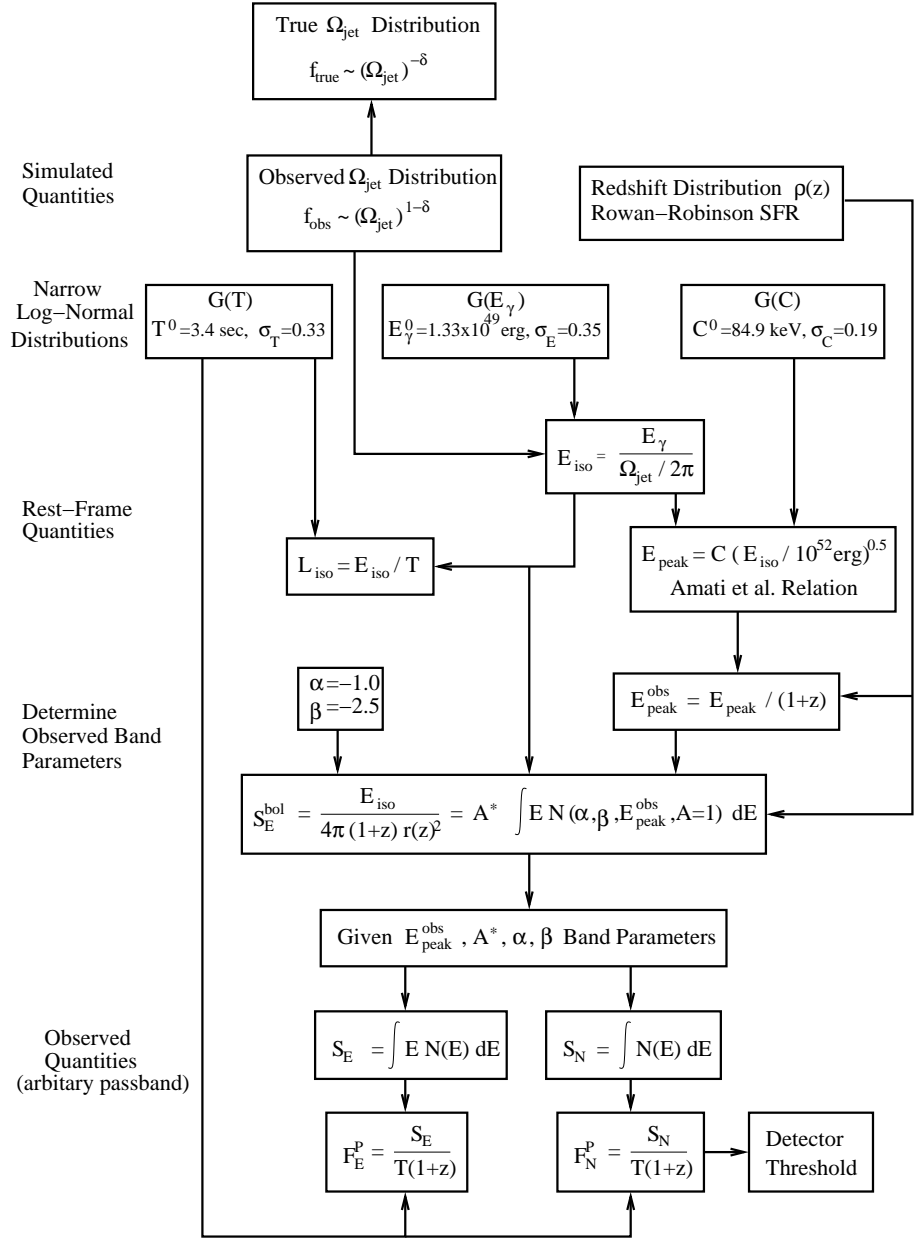}}
\end{center}  
\caption{Flowchart showing the logical sequence of the simulations
for the variable jet opening-angle model.}
\label{flowchart}
\end{figure}

%
\begin{figure}[ht]  
\begin{center}
\resizebox{5.8cm}{!}{\includegraphics{f7a.eps}}
\resizebox{5.8cm}{!}{\includegraphics{f7b.eps}}
\end{center}  
\begin{center}
\resizebox{5.8cm}{!}{\includegraphics{f7c.eps}}
\resizebox{5.8cm}{!}{\includegraphics{f7d.eps}}
\end{center}  
\begin{center}
\resizebox{5.8cm}{!}{\includegraphics{f7e.eps}}
\resizebox{5.8cm}{!}{\includegraphics{f7f.eps}}
\end{center}  
\caption{
{\it Left panels:} comparison of the best-fit smearing functions
$G(\egamma$), $G(C)$, and $G(T)$ and the cumulative distributions of
$\egamma$ (top), $\Delta \ep$ (middle), and $T_0$ (bottom),
respectively.
{\it Right panels:} $\egamma$ (top), the deviation $\Delta \ep$ in $\ep$ of
bursts from the $\eiso - \ep$ relation that we have adopted (middle),
and $T_0 \sim \fn/\se$ (bottom), as a function of redshift $z$. 
The smaller error bar represents the statistical error, while the larger
represents the total error.  Closed circles denote HETE-2 points, while
open circles denote \bsax points.  For more information see the text and
Table \ref{gauss_table}.
}
\label{gaussian_fit_plots}
\end{figure}

%
\begin{figure}[ht]  
\begin{center}
\rotatebox{270}{\resizebox{5.8in}{!}{\includegraphics{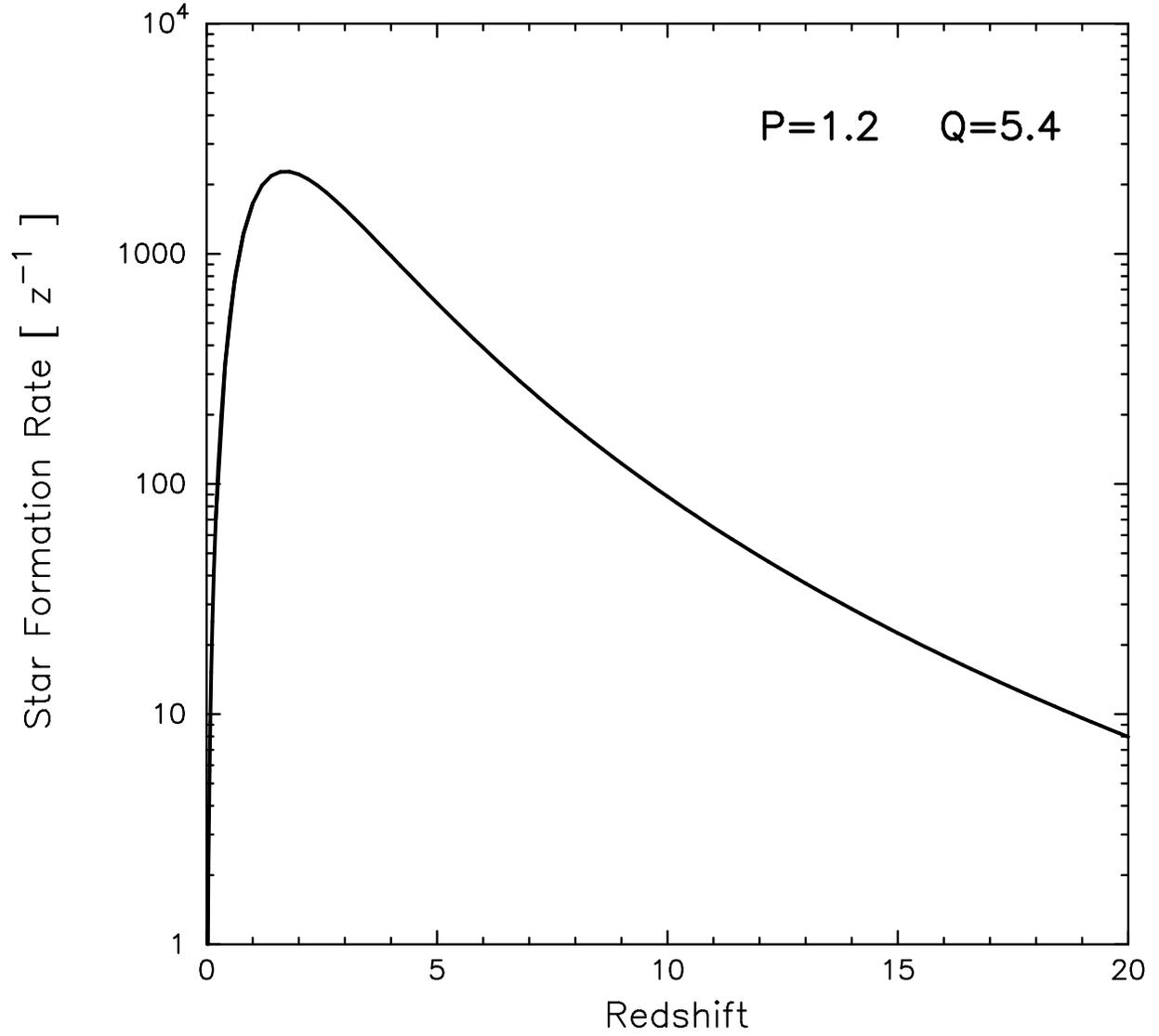}}}
\end{center}  
\caption{GRB rate as a function of redshift that we assume in this
paper.  The curve is the \cite{rr2001} function for the star-formation
rate [equation (3)], taking $P = 1.2$ and  $Q = 5.4$.}
\label{grb_rate}
\end{figure}

%
\begin{figure}[ht]
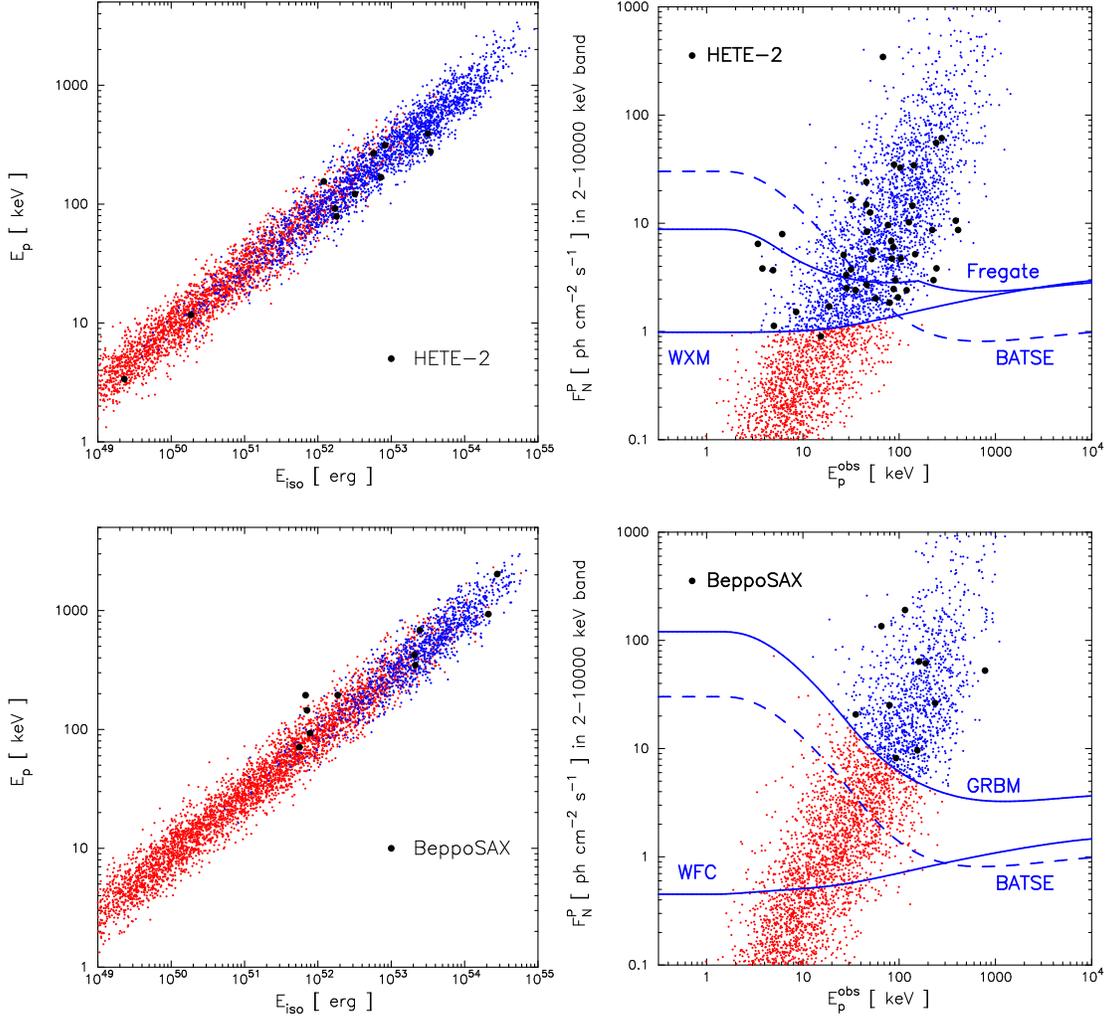
  
\begin{center}
\rotatebox{270}{\resizebox{6.5cm}{!}{\includegraphics[clip=]{f9a.eps}}}
\rotatebox{270}{\resizebox{6.5cm}{!}{\includegraphics[clip=]{f9b.eps}}}
\end{center}  
\begin{center}
\rotatebox{270}{\resizebox{6.5cm}{!}{\includegraphics[clip=]{f9c.eps}}}
\rotatebox{270}{\resizebox{6.5cm}{!}{\includegraphics[clip=]{f9d.eps}}}
\end{center}  
\caption{Comparison of the detectability of bursts by HETE-2 and \bsax
in the variable jet opening-angle model for $\delta = 2$.  Detected
bursts are shown in blue and non-detected bursts in red.  The left-hand
panels show bursts in the [$\eiso,\ep$]-plane detected by HETE-2 (upper
panel) and by \bsax (lower panel).  For each experiment, we overplot
the locations of the HETE-2 and \bsax bursts with known redshifts.  The
observed burst in the lower left-hand corner of the HETE-2 panel is XRF
020903, the most extreme burst in our sample.  The agreement between
the observed and predicted distributions of bursts is good.  The
right-hand panels show bursts in the [$\eop, \fn(2-10000\ {\rm
keV})$]-plane detected by HETE-2 (upper panel) and by \bsax (lower
panel).  For each experiment we show the sensitivity thresholds for
their respective instruments plotted as solid blue lines.  The BATSE
threshold is shown in both panels as a dashed blue line.  Again, the
agreement between the observed and predicted distributions of bursts is
good.  The left-hand panels exhibit the constant density of bursts per
logarithmic interval in $\eiso$ and $\ep$ given by the variable jet
opening-angle model for $\delta = 2$.
}
\label{uniform_jet_hete_bsax}
\end{figure}

%
\begin{figure}[ht]
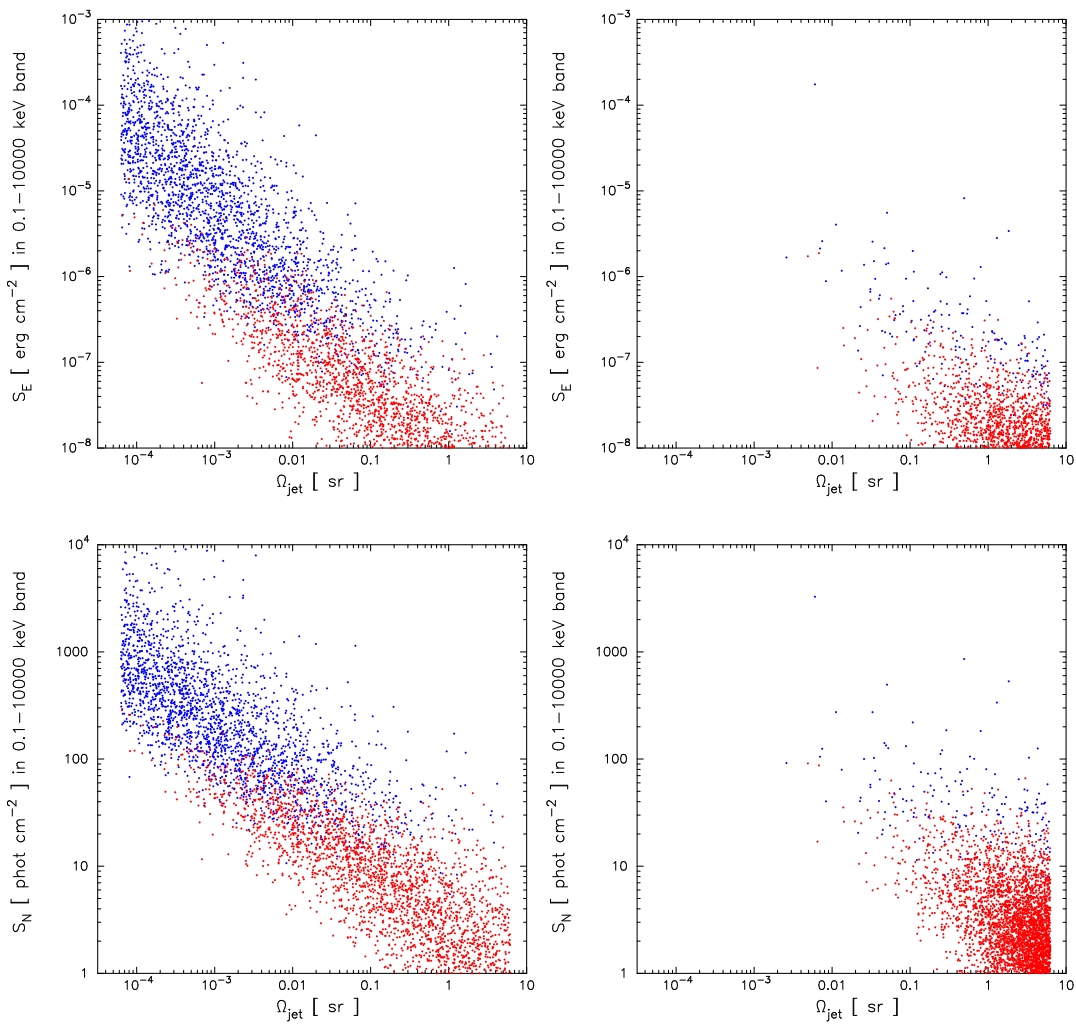
  
\begin{center}
\rotatebox{270}{\resizebox{6.5cm}{!}{\includegraphics[clip=]{f10a.eps}}}
\rotatebox{270}{\resizebox{6.5cm}{!}{\includegraphics[clip=]{f10b.eps}}}
\end{center}  
\begin{center}
\rotatebox{270}{\resizebox{6.5cm}{!}{\includegraphics[clip=]{f10c.eps}}}
\rotatebox{270}{\resizebox{6.5cm}{!}{\includegraphics[clip=]{f10d.eps}}}
\end{center}  
\caption{Scatter plots of $\se$ and $\sn$ versus $\ojet$.  The top
panels show the predicted distributions in the variable jet
opening-angle model for $\delta = 2$, while the bottom panels show the
power-law universal jet model pinned to the $\eiso$ value of XRF
020903 (see text).  Bursts detected by the WXM are shown in blue and
non-detected bursts in red.  The top panels exhibit the constant
density of bursts per logarithmic interval in $\se$, $\sn$, and $\ojet$
given by the variable jet opening-angle model for $\delta = 2$.  The
bottom panels exhibit the concentration of bursts at $\ojet \equiv
\oview \approx 2 \pi$ and the resulting preponderance of XRFs relative
to GRBs in the power-law universal jet model when it is pinned to
the $\eiso$ value of XRF 020903; i.e., when one attempts to extend the
model to include XRFs and X-ray-rich GRBs, as well as GRBs.}
\label{SE_SN_vs_omega}
\end{figure}

%
\begin{figure}[ht]  
\begin{center}
\rotatebox{270}{\resizebox{6.0cm}{!}{\includegraphics[clip=]{f11a.eps}}}
\rotatebox{270}{\resizebox{6.0cm}{!}{\includegraphics[clip=]{f11b.eps}}}
\end{center}  
\begin{center}
\rotatebox{270}{\resizebox{6.0cm}{!}{\includegraphics[clip=]{f11c.eps}}}
\rotatebox{270}{\resizebox{6.0cm}{!}{\includegraphics[clip=]{f11d.eps}}}
\end{center}  
\begin{center}
\rotatebox{270}{\resizebox{6.0cm}{!}{\includegraphics[clip=]{f11e.eps}}}
\rotatebox{270}{\resizebox{6.0cm}{!}{\includegraphics[clip=]{f11f.eps}}}
\end{center}  
\caption{Scatter plots of $\fe$ and $\fn$ versus $\ojet$.  The top
panels show the predicted distributions in the variable jet
opening-angle model for $\delta = 2$, while the middle and the bottom
panels show the power-law universal jet model pinned to the $\eiso$
values of XRF 020903 and GRB 980326, respectively (see text).  Bursts
detected by the WXM are shown in blue and non-detected bursts in red.
}
\label{FEP_FNP_vs_omega}
\end{figure}

%
\begin{figure}[ht]
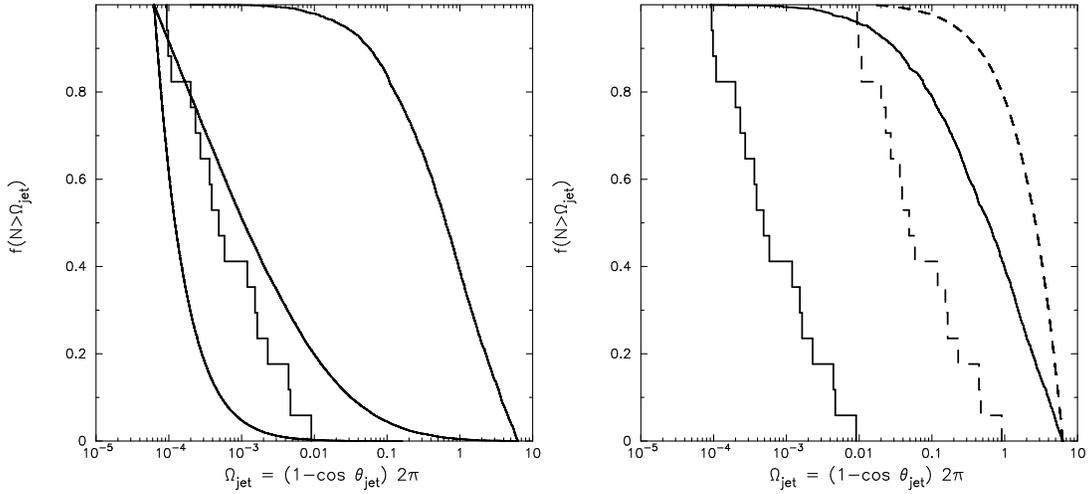
  
\begin{center}
\rotatebox{270}{\resizebox{6.5cm}{!}{\includegraphics[clip=]{f12a.eps}}}
\rotatebox{270}{\resizebox{6.5cm}{!}{\includegraphics[clip=]{f12b.eps}}}
\end{center}  
\caption{Observed and predicted cumulative distributions of $\ojet$. 
{\it Left panel:} Cumulative distributions of $\ojet$ predicted by the
variable jet opening-angle model for $\delta = 1$, $2$ and $3$ (curves
from right to left), compared to the observed cumulative distribution
of the values of $\ojet$ given in \cite{bloom2003} scaled downward by a
factor of $C_{\rm jet}$ = 95 (solid histogram).
{\it Right panel:} Cumulative $\ojet \equiv \oview$ distributions
predicted by the power-law universal jet model with the minimum value of $\eiso$
pinned to the value of $\eiso$ for XRF 020903 (solid curve) and to
the value of $\eiso$ for GRB 980326 (dashed curve).  Models are
compared with observed cumulative distribution of the values of $\ojet$
given in \cite{bloom2003} (dashed histogram) and the same scaled
downward by a factor of $C_{\rm jet}$ = 95 (solid histogram).  See
the text for more details.
}
\label{omega_cuml}
\end{figure}

%
\begin{figure}[ht]  
\begin{center}
\rotatebox{270}{\resizebox{5.5cm}{!}{\includegraphics[clip=]{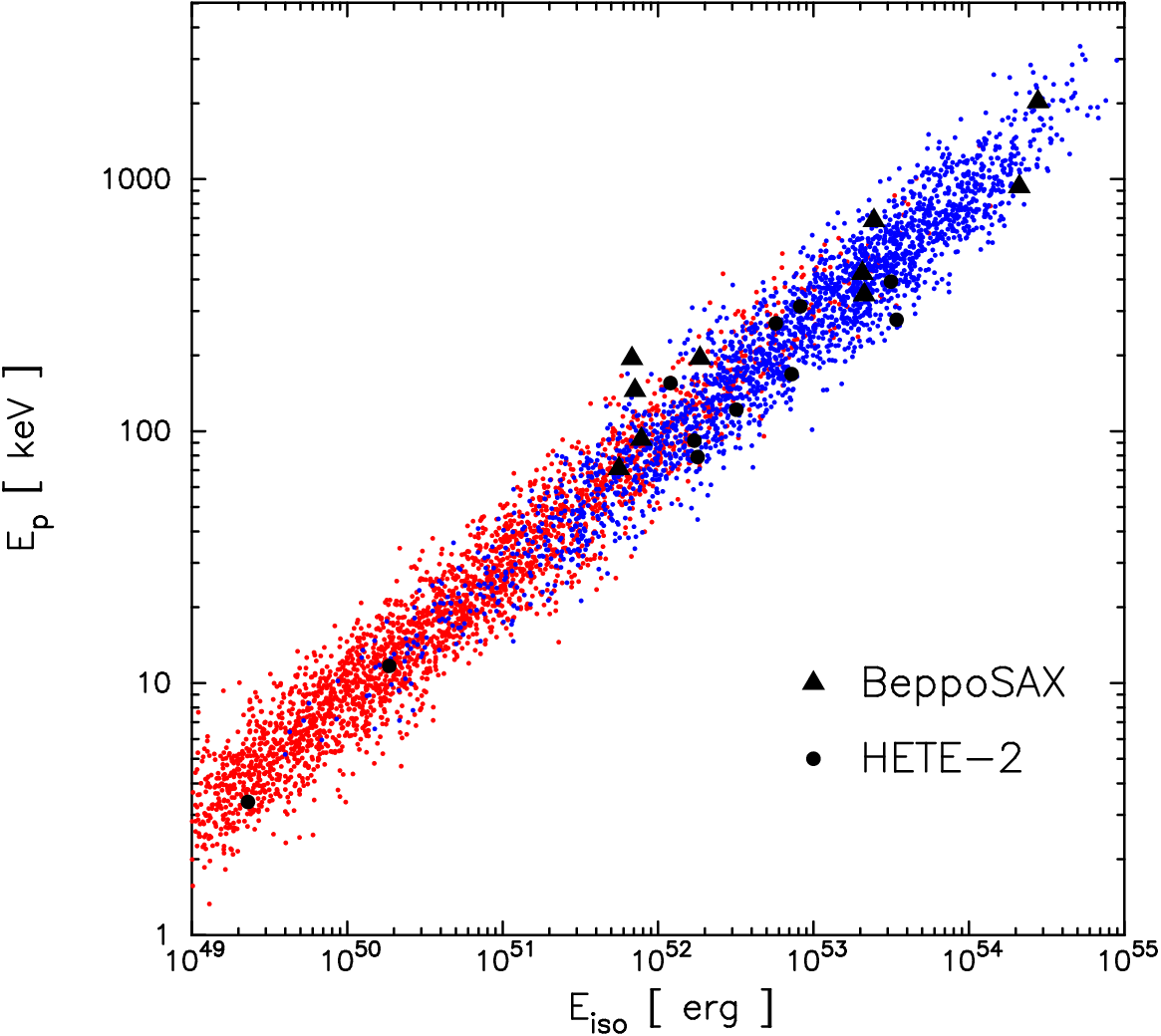}}}
\rotatebox{270}{\resizebox{5.5cm}{!}{\includegraphics[clip=]{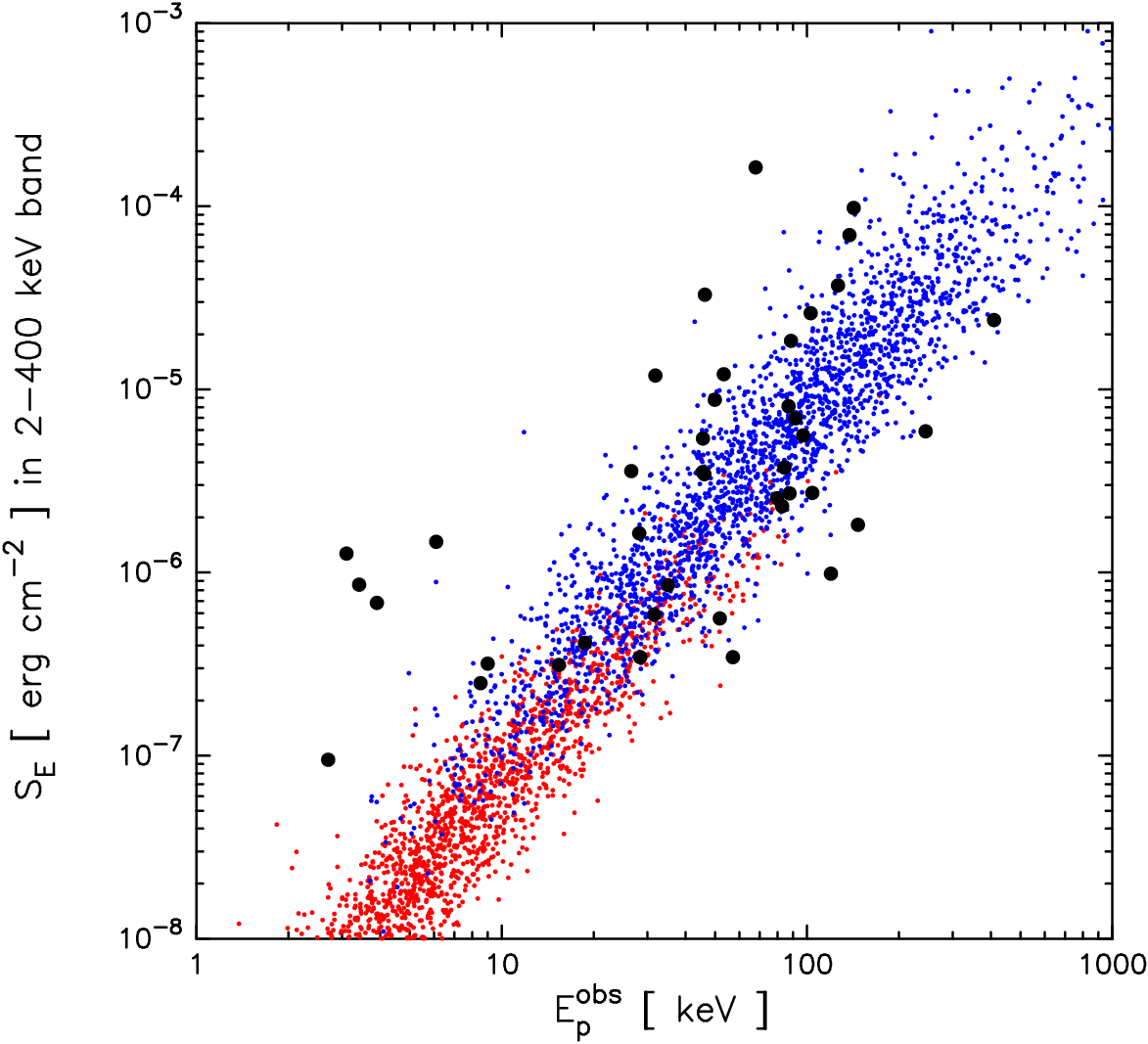}}}
\end{center}  
\begin{center}
\rotatebox{270}{\resizebox{5.5cm}{!}{\includegraphics[clip=]{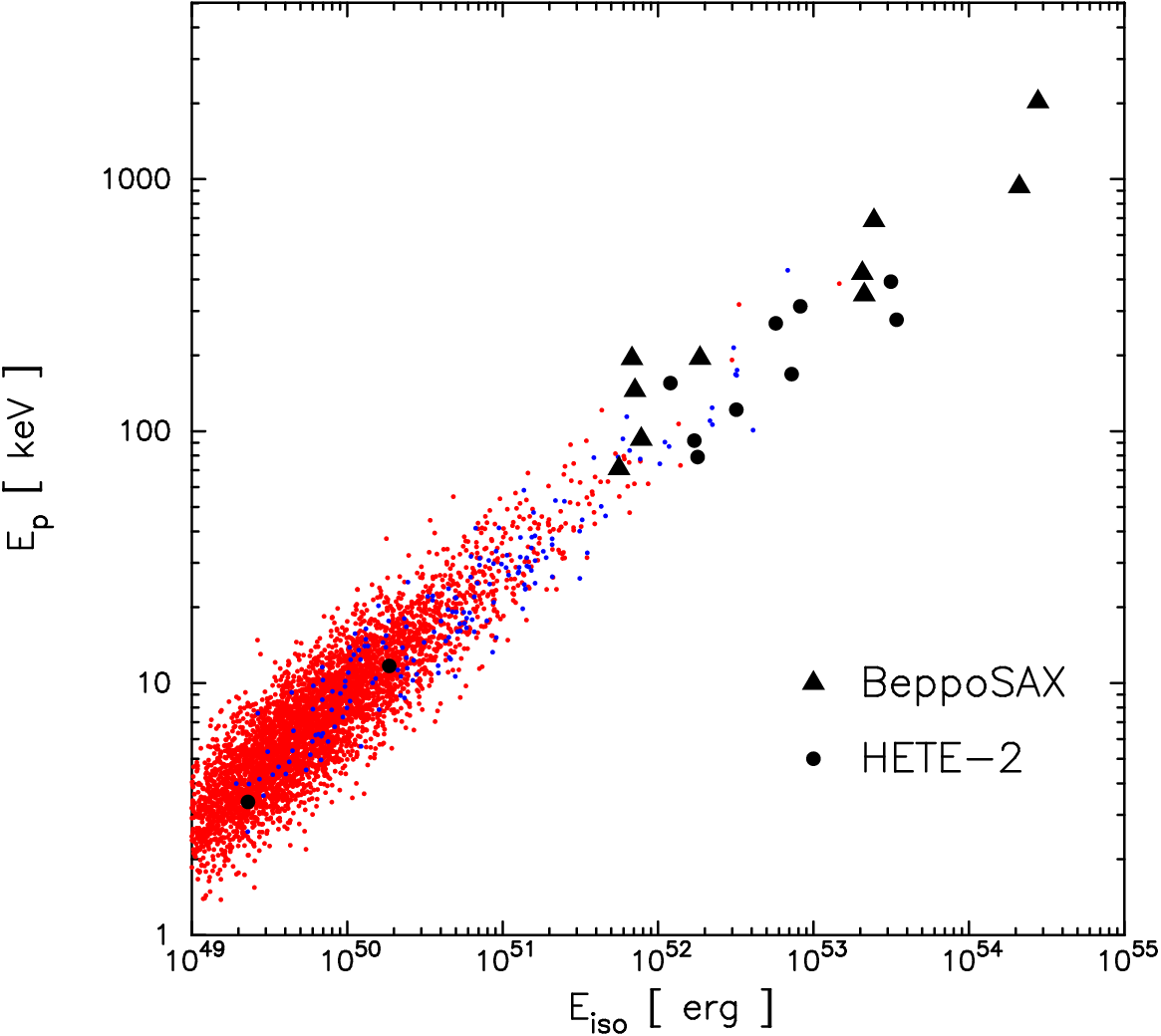}}}
\rotatebox{270}{\resizebox{5.5cm}{!}{\includegraphics[clip=]{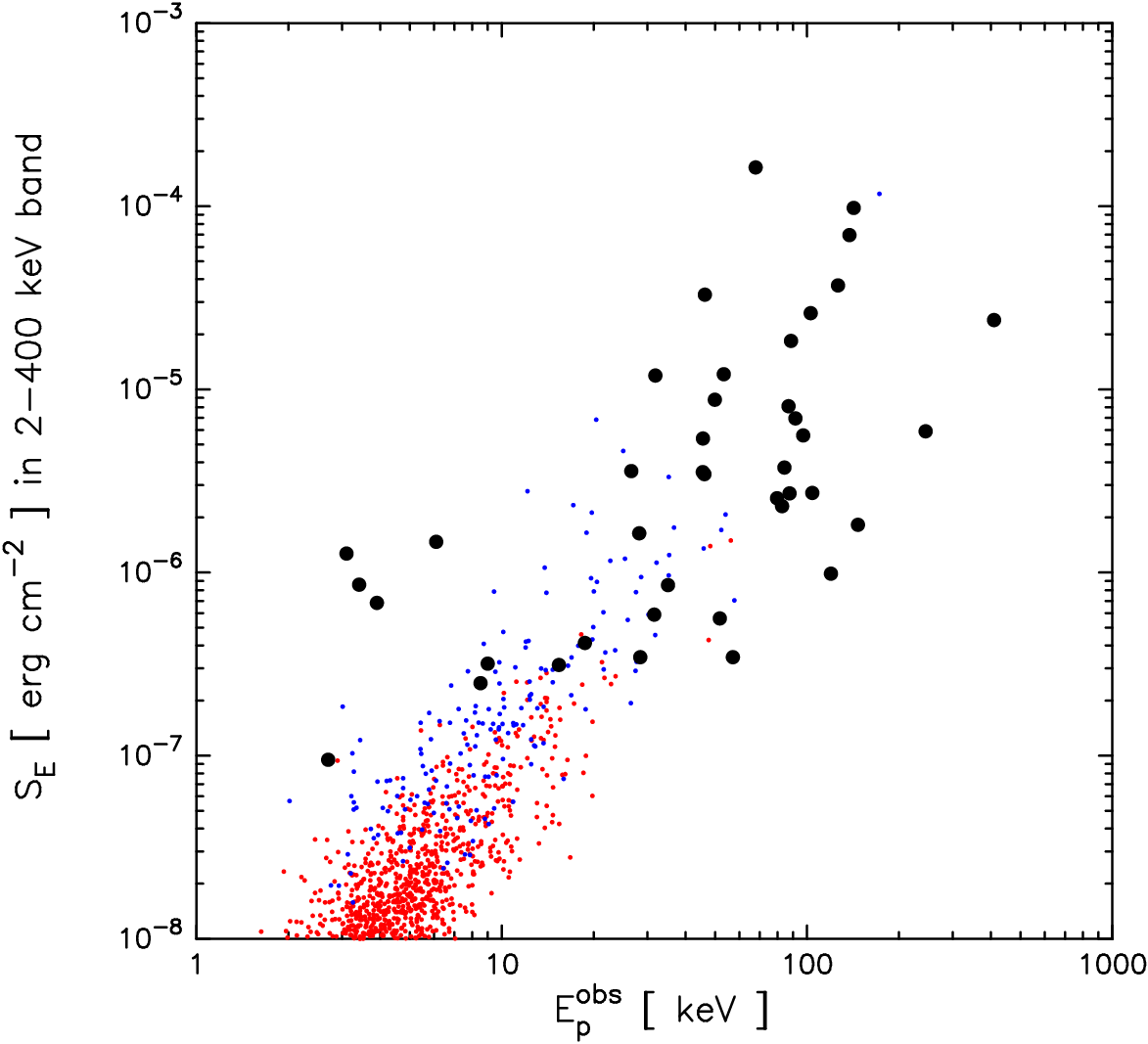}}}
\end{center}  
\begin{center}
\rotatebox{270}{\resizebox{5.5cm}{!}{\includegraphics[clip=]{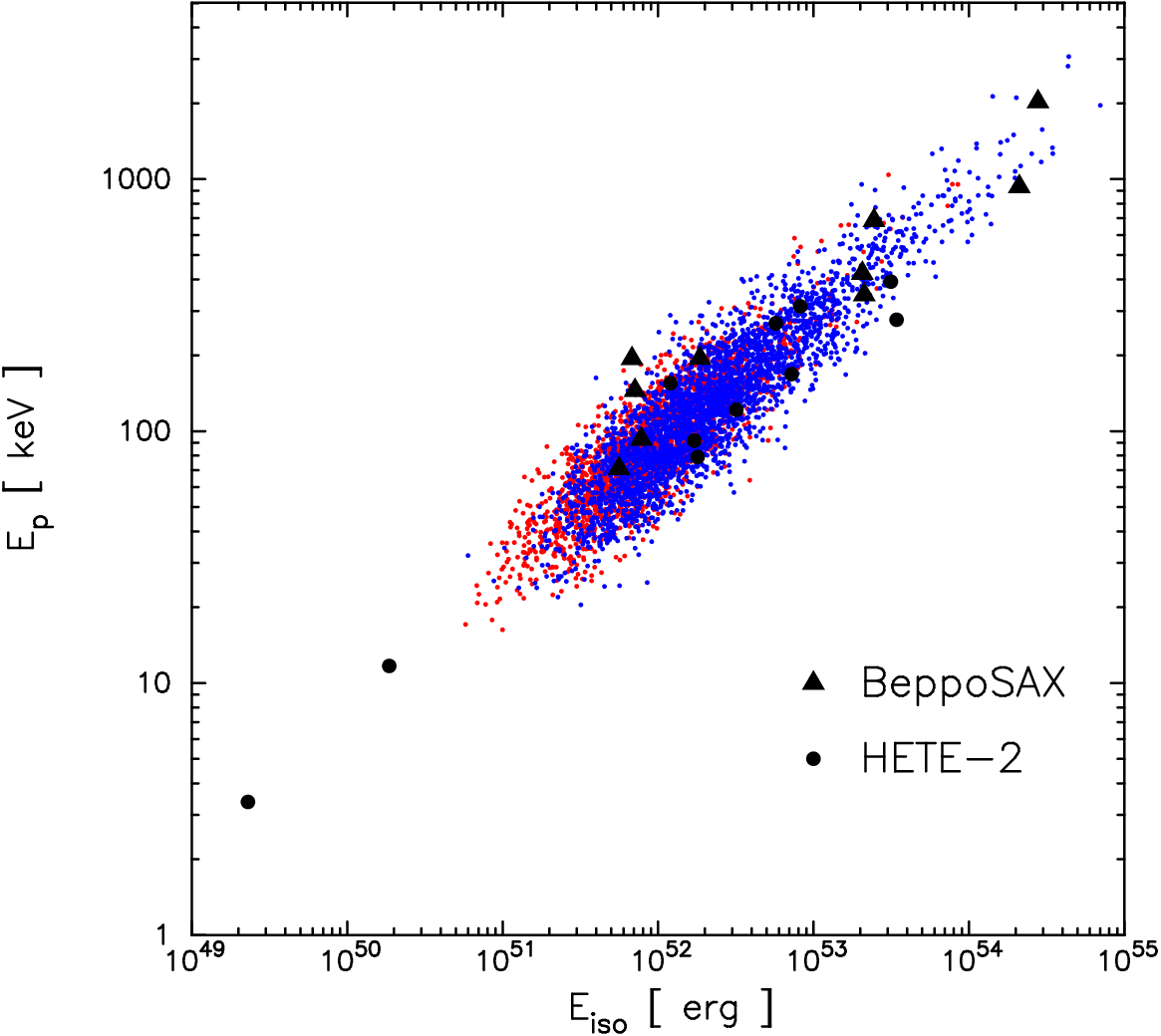}}}
\rotatebox{270}{\resizebox{5.5cm}{!}{\includegraphics[clip=]{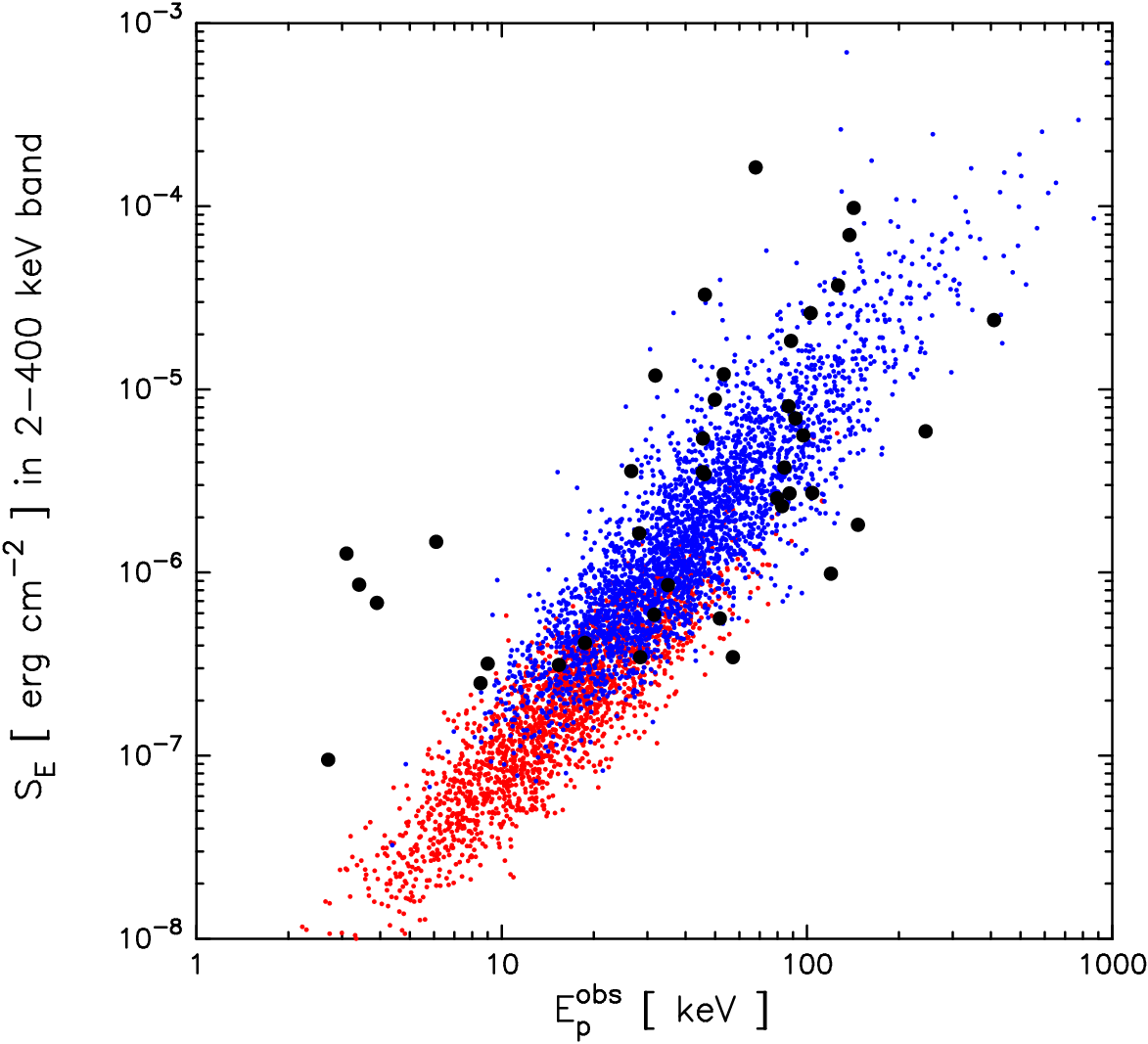}}}
\end{center}  
\caption{Scatter plots of $\eiso$ versus $\ep$ (left column) and $\ep$
versus $\se$ (right column).  The top panels show the predicted
distributions in the variable jet opening-angle model for $\delta = 2$,
while the middle and the bottom panels show the power-law universal jet model
pinned to the $\eiso$ values of XRF 020903 and GRB 980326,
respectively (see text).  Bursts detected by the WXM are shown in blue
and non-detected bursts in red.  In the left column, the triangles and
circles respectively show the locations of the \bsax and HETE-2 bursts
with known redshifts.  In the right column, the black circles show the
locations of all HETE-2 bursts for which joint fits to WXM and FREGATE
spectral data have been carried out \citep{sakamoto2003b}.
}
\label{3models_scatter}
\end{figure}

%
\begin{figure}[ht]  
\begin{center}
\rotatebox{270}{\resizebox{5.5cm}{!}{\includegraphics[clip=]{f14a.eps}}}
\rotatebox{270}{\resizebox{5.5cm}{!}{\includegraphics[clip=]{f14b.eps}}}
\end{center}  
\begin{center}
\rotatebox{270}{\resizebox{5.5cm}{!}{\includegraphics[clip=]{f14c.eps}}}
\rotatebox{270}{\resizebox{5.5cm}{!}{\includegraphics[clip=]{f14d.eps}}}
\end{center}  
\begin{center}
\rotatebox{270}{\resizebox{5.5cm}{!}{\includegraphics[clip=]{f14e.eps}}}
\rotatebox{270}{\resizebox{5.5cm}{!}{\includegraphics[clip=]{f14f.eps}}}
\end{center}  
\caption{Comparison of the observed and predicted cumulative
distributions of $\eiso$ (upper row) and $\ep$ (middle row) for HETE-2
and \bsax bursts with known redshifts; and $\se$ and $\eop$ for all
HETE-2 bursts (lower row).  The solid histograms are the observed
cumulative distributions.  The blue curves are the cumulative
distributions predicted by the variable jet opening-angle model for
$\delta = 2$.  The solid red curves are the cumulative distributions
predicted by the power-law universal jet model pinned at the $\eiso$ value of
XRF 020903; i.e., when one attempts to extend the model to include XRFs
and X-ray-rich GRBs, as well as GRBs.  The dashed red curves are the
cumulative distributions predicted by the power-law universal jet model
pinned at the $\eiso$ value of GRB 980326; i.e., when one fits the
model only to GRBs.  
}
\label{fab_four}
\end{figure}

%
\begin{figure}[ht]
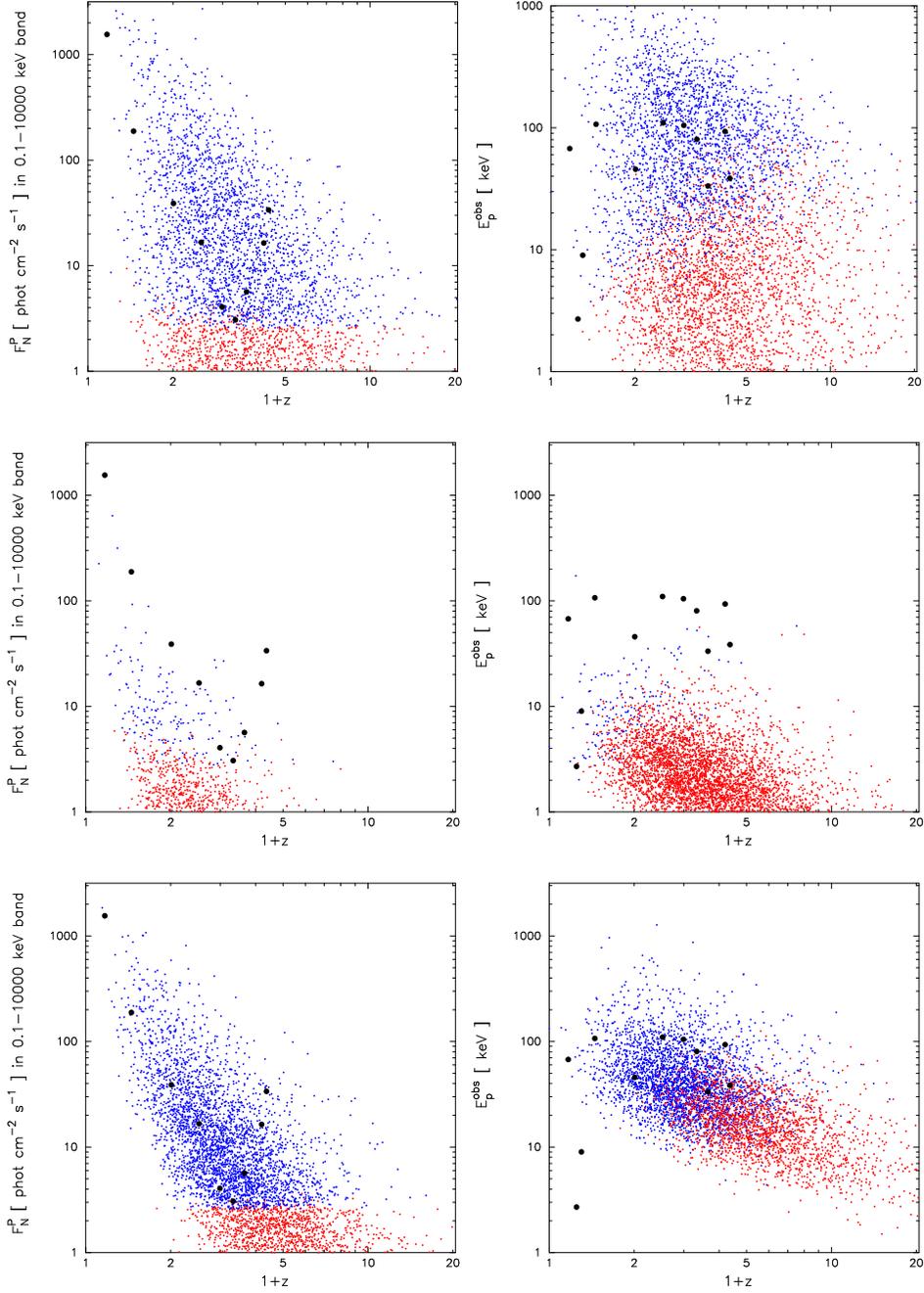
  
\begin{center}
\rotatebox{270}{\resizebox{5.5cm}{!}{\includegraphics[clip=]{f15a.eps}}}
\rotatebox{270}{\resizebox{5.5cm}{!}{\includegraphics[clip=]{f15b.eps}}}
\end{center}  
\begin{center}
\rotatebox{270}{\resizebox{5.5cm}{!}{\includegraphics[clip=]{f15c.eps}}}
\rotatebox{270}{\resizebox{5.5cm}{!}{\includegraphics[clip=]{f15d.eps}}}
\end{center}  
\begin{center}
\rotatebox{270}{\resizebox{5.5cm}{!}{\includegraphics[clip=]{f15e.eps}}}
\rotatebox{270}{\resizebox{5.5cm}{!}{\includegraphics[clip=]{f15f.eps}}}
\end{center}  
\caption{Scatter plots of $\fn$ (left column) and $\eop$ (right column)
as a function of redshift.  The top row shows the distributions of
bursts predicted by the variable jet opening-angle model for $\delta =
2$.  The middle row shows the distributions of bursts predicted by the
power-law universal jet model pinned to the $\eiso$ value for XRF 020903,
while the bottom row shows the distributions of bursts predicted by the
power-law universal jet model pinned to the $\eiso$ value for GRB 980326. 
Bursts detected by the WXM are shown in blue and non-detected bursts in
red.  The filled black circles show the positions of the \bsax and
HETE-2 bursts with known redshifts.
}
\label{FNP_EPO_z1_scatter}
\end{figure}

%
\begin{figure}[ht]
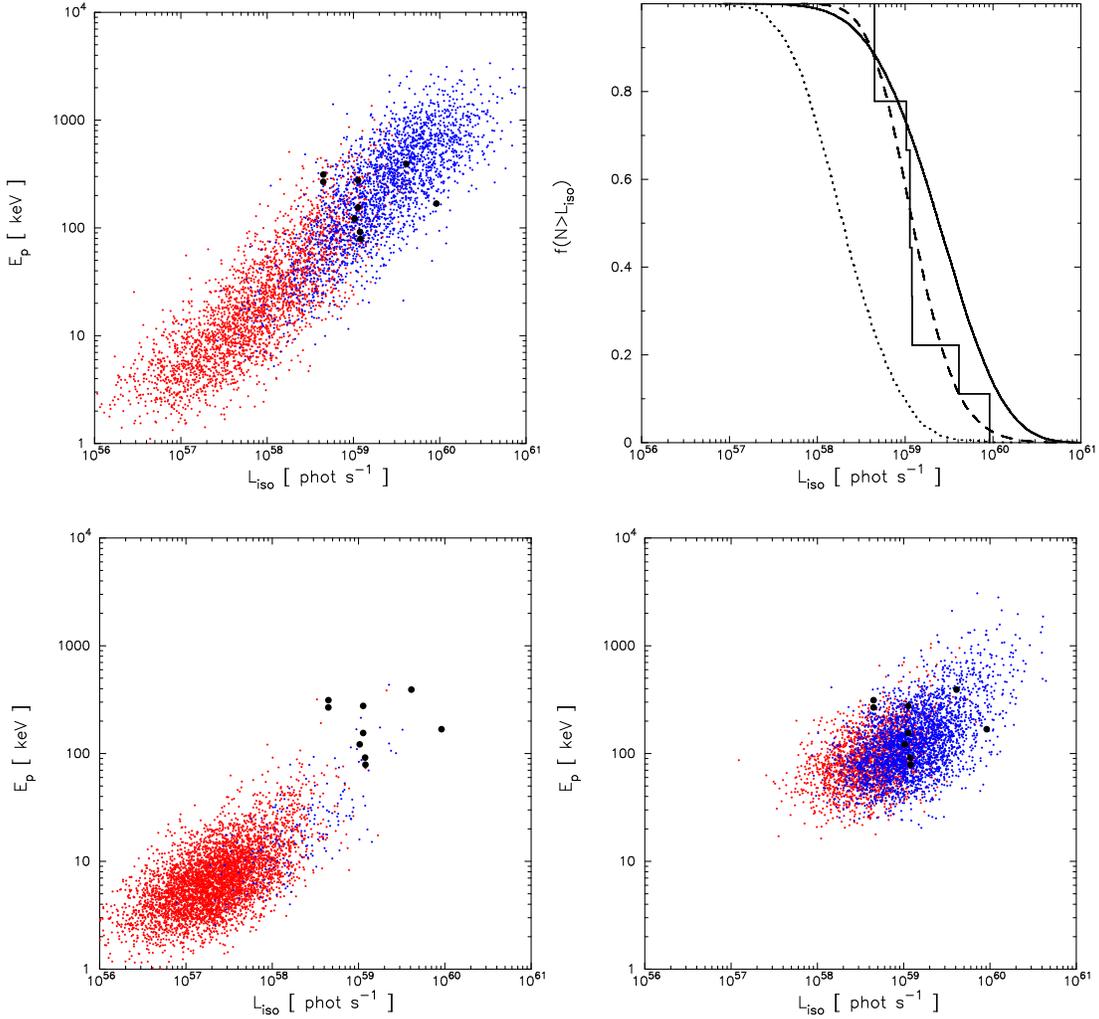
  
\begin{center}
\rotatebox{270}{\resizebox{6.5cm}{!}{\includegraphics[clip=]{f16a.eps}}}
\rotatebox{270}{\resizebox{6.5cm}{!}{\includegraphics[clip=]{f16b.eps}}}
\end{center}  
\begin{center}
\rotatebox{270}{\resizebox{6.5cm}{!}{\includegraphics[clip=]{f16c.eps}}}
\rotatebox{270}{\resizebox{6.5cm}{!}{\includegraphics[clip=]{f16d.eps}}}
\end{center}  
\caption{Scatter plots of $\liso$ versus $\ep$ and comparison of
observed and predicted cumulative distributions of $\liso$.   
{\it Upper left}: Distribution of bursts predicted by the variable jet
opening-angle model for $\delta = 2$. Bursts detected by the WXM are
shown in blue and non-detected bursts in red. The filled black circles
show the positions of the HETE-2 bursts with known redshifts. 
{\it Upper right}: Observed cumulative distribution of $\liso$ for
HETE-2 bursts with known redshifts (histogram) compared against the
cumulative $\liso$ distributions predicted by the variable jet
opening-angle model for $\delta = 2$ (solid curve) and by the power-law
universal jet model pinned at the $\eiso$ value for XRF 020903
(dotted curve) and for GRB 980326 (dashed curve).
{\it Lower left}: Distribution of bursts predicted by the power-law
universal jet model pinned at the $\eiso$ value for XRF 020903. 
{\it Lower right}: Distribution of bursts predicted by the power-law
universal jet model pinned at the $\eiso$ value for GRB 980326.
}
\label{Lgamma_scatter_cuml}
\end{figure}

%
\begin{figure}[ht]
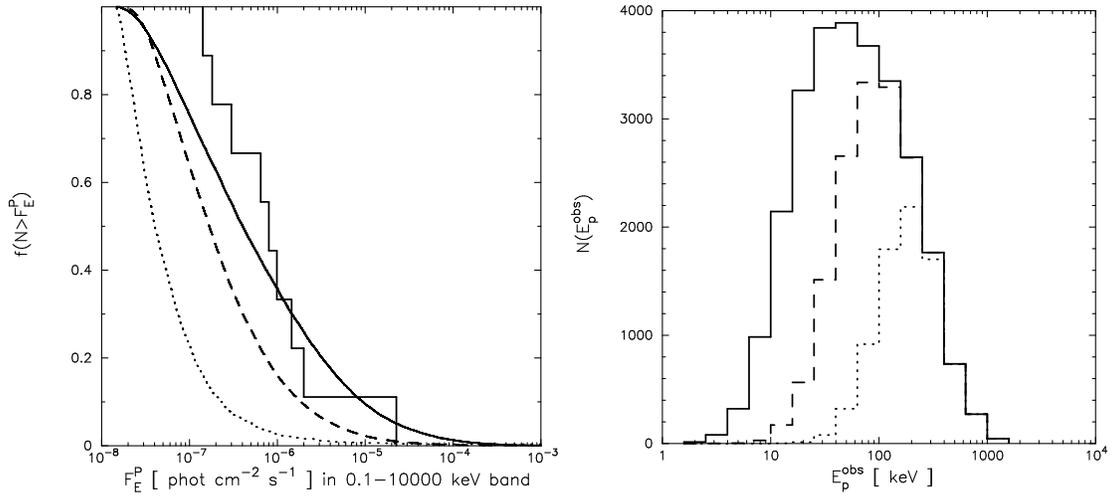
  
\begin{center}
\rotatebox{270}{\resizebox{6.5cm}{!}{\includegraphics[clip=]{f17a.eps}}}
\rotatebox{270}{\resizebox{6.5cm}{!}{\includegraphics[clip=]{f17b.eps}}}
\end{center}  
\caption{
{\it Left panel}: Comparison of the observed cumulative distribution of
$\fe$ for HETE-2 bursts (histogram) with the cumulative $\fe$
distributions predicted by the variable jet opening-angle model for
$\delta = 2$ (solid curve) and by the power-law universal jet model
pinned at the $\eiso$ value for XRF 020903 (dotted curve) and at
the $\eiso$ value for GRB 980326 (dashed curve). 
{\it Right panel}: Differential distribution of $\eop$ predicted by the
variable jet opening-angle model for $\delta = 2$ for bursts with $\fe
> 10^{-8}$ (solid histogram), $10^{-7}$ (dashed histogram), and $
10^{-6}$ erg cm$^{-2}$ s$^{-1}$ (dotted histogram).
}
\label{FEP_dists}
\end{figure}

\clearpage

%
\begin{figure}[ht]  
\begin{center}
\rotatebox{270}{\resizebox{11.0cm}{!}{\includegraphics{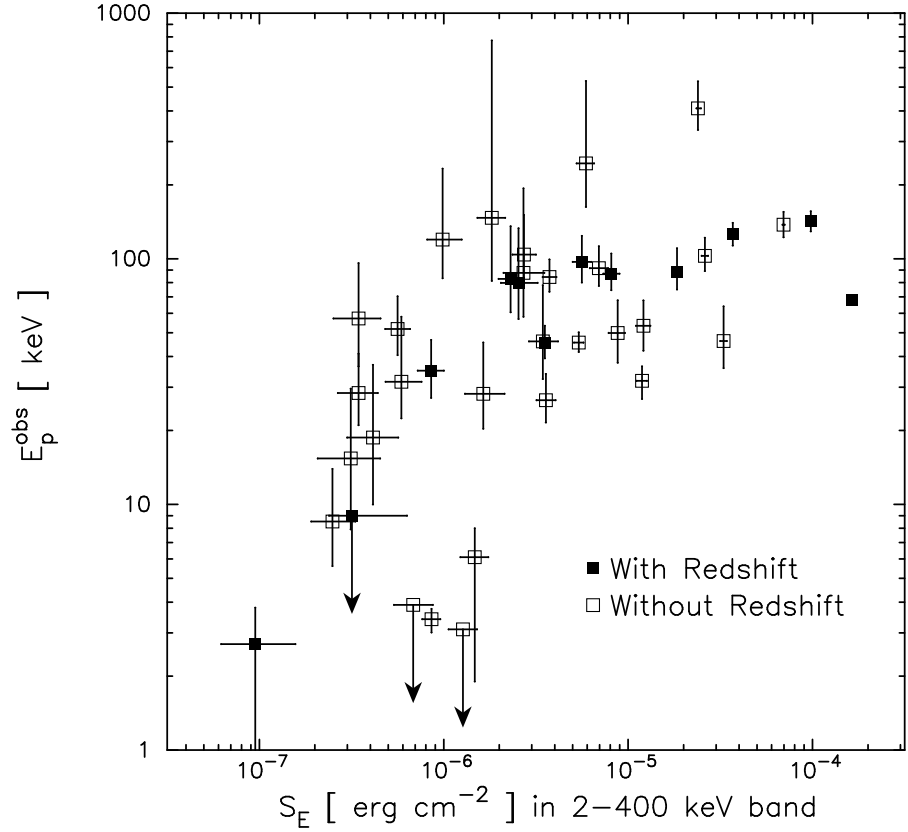}}}
\end{center}  
\caption{
Distribution of HETE-2 bursts in the [$S (2-400 {\rm keV}), E^{\rm
obs}_{\rm peak}$]-plane, showing the bursts with redshift
determinations (solid squares) and those without (open squares).  The
two events with known redshifts in the lower left-hand corner of the
figure are XRF 020903 and XRF 030723.  From \cite{sakamoto2003b}.
}
\label{hete_Epobs_flu_by_z}
\end{figure}

\clearpage

%
\begin{figure}[ht]
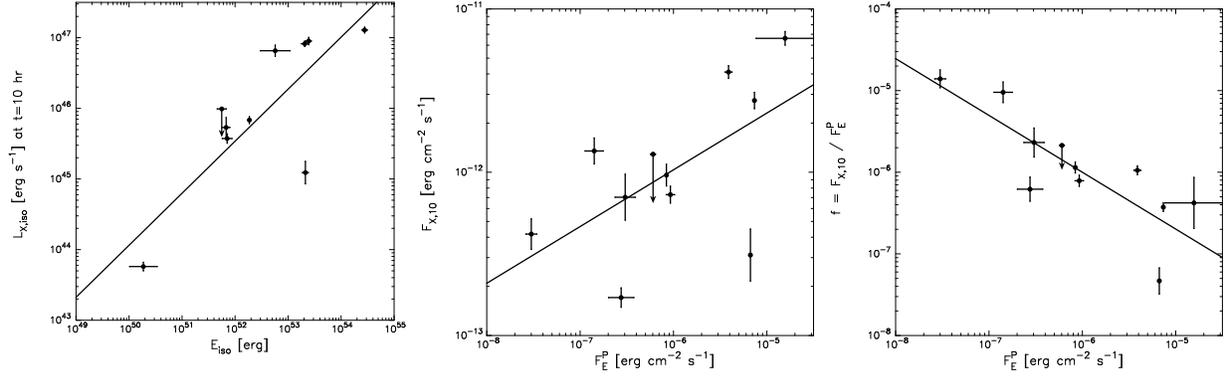
  
\begin{center}
\resizebox{5.3cm}{!}{\rotatebox{270}{\includegraphics{f19a.eps}}}
\resizebox{5.3cm}{!}{\rotatebox{270}{\includegraphics{f19b.eps}}}
\resizebox{5.3cm}{!}{\rotatebox{270}{\includegraphics{f19c.eps}}}
\end{center}
\caption{
{\it Left panel:} Correlation between the isotropic-equivalent burst
energy ($\eiso$) and the X-ray afterglow luminosity ($\lx$) at 10 hours
after the burst from \cite{berger2003}.  The slope of the best-fit line
is $0.74 \pm 0.17$ (68\% confidence level).
{\it Middle panel:} Correlation between the X-ray afterglow flux at 10
hours after the burst ($F_{\rm x,10}$) and the peak energy flux
($\fe$).  The slope of the best-fit line is $0.35 \pm 0.14$ (68\%
confidence level).
{\it Right panel:} The ratio $f = F_{\rm x,10} / \fe$ as a function of
$\fe$.  The slope of the best-fit line is $-0.70 \pm 0.15$ (68\%
confidence level).
After Lamb et al. (2005, in preparation).
}
\label{LX_Eiso}
\end{figure}

\clearpage


\clearpage


\begin{thebibliography}{999}

\bibitem[Amati et al.(2002)]{amati2002} 
	Amati, L., et al. 2002, \aap, 390, 81
\bibitem[Atteia et al.(2003)]{atteia2003}
	Atteia, J-L, et al. 2003, in AIP Conf. Proc. 662, Gamma-Ray Burst 
	and Afterglow Astronomy 2001, ed. G. R. Ricker \&
	R. K. Vanderspek (New York: AIP), 17
\bibitem[Band, et al.(1993)]{band1993} 
	Band, D. L., et al. 1993, ApJ, 413, 281
\bibitem[Band(2003)]{band2003} 
	Band, D. L. 2003, ApJ, 588, 945
\bibitem[Barraud et al.(2003)]{barraud2003}
	Barraud, C., et al. 2003, A\&A, 400, 1021
\bibitem[Berger et al.(2003a)]{berger2003}
	Berger, E., Kulkarni, S. R., Frail, D. A., \& Soderberg, A. M.
	2003a, ApJ, 590, 379
\bibitem[Berger et al.(2003b)]{berger2003b}
	Berger, E., et al. 2003b, Nature, 426, 154
\bibitem[Berger et al.(2003c)]{berger2003c}
	Berger, E., et al. 2003c, ApJ, 599, 408
\bibitem[Bloom, Frail \& Kulkarni(2003)]{bloom2003} 
	Bloom, J., Frail, D. A., \& Kulkarni, S. R. 2003, ApJ, 588, 945
\bibitem[Butler et al.(2003a)]{butler2003a}
	Butler, N., et al. 2003a, GCN Circular 2328
\bibitem[Butler et al.(2003b)]{butler2003b}
	Butler, N., et al. 2003b, GCN Circular 2347	
\bibitem[Coburn \& Boggs(2003)]{coburn2003}
	Coburn, W., \& Boggs, S. E. 2003, Nature, 423, 415
\bibitem[Costa \& Frontera(2003)]{costa2003}
	Costa, E., \& Frontera, F. 2003, private communication
\bibitem[Dermer et al.(1999)]{dermer1999} 
	Dermer, C. D., Chiang, J., and B$\ddot{\rm o}$ttcher 
	1999, \apj, 513, 656 
\bibitem[Dermer and Mitman(2003)]{dermer2003} 
    Dermer, C. D., and Mitman, K. E. 2003, in ASP Conf. Ser. 312, Third
    Rome Workshop on Gamma-Ray Bursts in the Afterglow Era, ed. M. Feroci,
    F. Frontera, N. Masetti \& L. Piro (ASP: San Francisco), 301
\bibitem[Donaghy, Graziani \& Lamb(2004a)]{donaghy2004a}
	Donaghy, T. Q., Lamb, D. Q., \& Graziani, C., 2004a, ApJ, submitted
\bibitem[Donaghy, Graziani \& Lamb(2004b)]{donaghy2003b}
	Donaghy, T. Q., Lamb, D. Q., \& Graziani, C. 2004b, in AIP Conf.
	Proc. 727, Gamma-Ray Bursts: 30 Years of Discovery, ed. E. E.
	Fenimore \& M. Galassi (Melville: AIP), 47
\bibitem[Donaghy(2004)]{donaghy2004c}
	Donaghy, T. Q. 2004, in preparation
\bibitem[Dullighan et al.(2003)]{dullighan2003}
	Dullighan, A., Butler, N., Vanderspek, E., J. Villasenor, J.,
	and Ricker, G. 2003, GCN Circular 2336
\bibitem[Fendt \& Ouyed(2003)]{fendt2003}
	Fendt, C. \& Ouyed, R. 2004, ApJ, 608, 378
\bibitem[Fenimore \& Ramirez-Ruiz(2000)]{fenimore2000}
	Fenimore, E. E., \& Ramirez-Ruiz, E. 2000, submitted to ApJ
	(astro-ph/0004176)
\bibitem[Fox et al.(2003)]{fox2003c}
	Fox, D. W., et al. 2003, GCN Circular 2323
\bibitem[Frail et al.(2001)]{frail2001}
	Frail, D., et al. 2001, ApJ, 562, L55
\bibitem[Heise et al.(2000)]{heise2000}  
	Heise, J., in't Zand, J., Kippen, R. M., \& Woods, P. M., in
	Proc. 2nd Rome Workshop:  Gamma-Ray Bursts in the Afterglow
	Era, eds. E. Costa, F. Frontera, J. Hjorth (Berlin:
	Springer-Verlag), 16
\bibitem[Huang, Dai \& Lu(2002)]{huang2002}
	Huang, Y. F., Dai, Z. G., \& Lu, T. 2002, MNRAS, 332, 735 
\bibitem[Kawai et al.(2003)]{kawai2003}
	Kawai, N., et al. 2003, in AIP Conf. Proc. 662, Gamma-Ray Burst 
	and Afterglow Astronomy 2001, ed. G. R. Ricker \&
	R. K. Vanderspek (New York: AIP), 25
\bibitem[Kippen et al.(2002)]{kippen2002}
	Kippen, R. M., Woods, P. M., Heise, J., in't Zand, J., Briggs,
	M.S., \& Preece, R. D. 2003, in AIP Conf. Proc. 662, Gamma-Ray Burst 
	and Afterglow Astronomy, ed. G. R. Ricker \& R. K.
	Vanderspek (New York: AIP), 244
\bibitem[Kumar \& Panaitescu(2003)]{kumar2003}
	Kumar, P., \& Panaitescu, A. 2003, MNRAS, 346, 905
\bibitem[Lamb(1999)]{lamb1999} 
	Lamb, D. Q. 1999, A\&A, 138, 607 
\bibitem[Lamb(2000)]{lamb2000}
	Lamb, D. Q. 2000, Physics Reports, 333-334, 505
\bibitem[Lamb, Donaghy, \& Graziani(2004a)]{lamb2004a} 
	Lamb, D. Q., Donaghy, T. Q., and Graziani, C. 2004a, 
	New Astronomy, 48, 423
\bibitem[Lamb Donaghy, \& Graziani(2004b)]{lamb2004b} 
	Lamb, D. Q., Donaghy, T. Q., and Graziani, C. 2004b, in
	Cosmic Explosions in Three Dimensions, ed. P. H\"oflich, P.
	Kumar, \& J. C. Wheeler (Cambridge: Cambridge Univ. Press), 327
\bibitem[Lamb Donaghy, \& Graziani(2004c)]{lamb2004c} 
	Lamb, D. Q., Donaghy, T. Q., and Graziani, C. 2004c, in AIP Conf.
	Proc. 727, Gamma-Ray Bursts: 30 Years of Discovery, ed. E. E.
	Fenimore \& M. Galassi (Melville: AIP), 19
\bibitem[Lamb et al.(2004)]{lamb2003} 
	Lamb, D. Q., et al. 2004d, ApJ, submitted
\bibitem[Liang, Dai, \& Wu(2004)]{liang2004}
	Liang, E. W., Dai, Z. G., \& Wu, X. F. 2004, ApJ, 606, L29
\bibitem[Lloyd-Ronning, Fryer \& Ramirez-Ruiz(2002)]{lloyd-ronning2002}
	Lloyd-Ronning, N., Fryer, C., \& Ramirez-Ruiz, E. 2002, ApJ,
	574, 554
\bibitem[Lloyd-Ronning, Petrosian \& Mallozzi(2000)]{lloyd-ronning2000}
	Lloyd-Ronning, N., Petrosian, V., \& Mallozzi, R. S. 2000, ApJ,
	534, 227
\bibitem[Lloyd-Ronning \& Zhang(2004)]{lloyd-ronning2004}
	Lloyd-Ronning, N. \& Zhang, B. 2004, ApJ, 613, 477
\bibitem[M\'esz\'aros, Ramirez-Ruiz, Rees \& Zhang(2002)]{meszaros2002}
	M\'esz\'aros, P., Ramirez-Ruiz, E., Rees, M. J., \& Zhang, B.
	2002, \apj, 578, 812
\bibitem[Mochkovitch et al.(2003)]{mochkovitch2003}
	Mochkovitch, R., Daigne, F., Barraud, C., \& Atteia, J. L.
	2004, in ASP Conf. Ser. 312, Third Rome Workshop on Gamma-Ray
	Bursts, ed. M. Feroci, F. Frontera, N. Masetti, \& L. Piro 
	(San Francisco: ASP), 381
\bibitem[Panaitescu \& Kumar(2001)]{panaitescu2001}
	Panaitescu, A., \& Kumar, P. 2001, ApJ, 560, L49
\bibitem[Panaitescu \& Kumar(2003)]{panaitescu2003}
	Panaitescu, A., \& Kumar, P. 2003, ApJ, 593, 290
\bibitem[Perna, Sari \& Frail(2003)]{perna2003}
	Perna, R., Sari, R., \& Frail, D. 2003, ApJ, 594, 379
\bibitem[Preece et al.(2000)]{preece2000}
	Preece, R., et al. 2000, ApJS, 126, 19
\bibitem[Prigozhin et al.(2003)]{prigozhin2003}
	Prigozhin, G., et al. 2003, GCN Circular 2313
\bibitem[Proga et al.(2003)]{proga2003}
	Proga, D., MacFadyen, A. I., Armitage, P. J., \& Begelman, M. C. 
	2003, ApJ, 599, L5
\bibitem[van Putten \& Levinson(2002)]{putten2002}
	van Putten, M. H. P. M., \& Levinson, A. 2002, Class. Quantum
	Grav., 19, 1309
\bibitem[Ramirez-Ruiz \& Lloyd-Ronning(2002)]{ramirez-ruiz2002}
	Ramirez-Ruiz, E. \& Lloyd-Ronning, N. 2002, New Astronomy, 7, 197
\bibitem[Reichart et al.(2001a)]{reichart2001}
	Reichart, D. E., et al. 2001a, ApJ, 522, 57
\bibitem[Reichart \& Lamb(2001)]{reichart2001b}
	Reichart, D. E., \& Lamb, D. Q. 2001, in AIP Conf. Proc. 586, 
	20th Texas Symposium on Relativistic Astrophysics, ed. 
	J. C. Wheeler \& H. Martel (New York: AIP), 938
\bibitem[Rhoads(1999)]{rhoads1999}
	Rhoads, J. E. 1999, ApJ, 525, 737
\bibitem[Ricker et al.(2003)]{ricker2003}
	Ricker, G. R. et al. 2003, in AIP Conf. Proc. 662, Gamma-Ray Burst
	and Afterglow Astronomy 2001, ed. G. R. Ricker \&
	R. K. Vanderspek (New York: AIP), 3
\bibitem[Rol et al.(2003)]{rol2003}
	Rol, E., et al. 2003, A\&A, 405, L23
\bibitem[Rossi, Lazzati \& Rees(2002)]{rossi2002} 
	Rossi, E., Lazzati, D, \& Rees, M. J. 2002, MNRAS 332, 945
\bibitem[Rowan-Robinson(2001)]{rr2001} 
	Rowan-Robinson, M. 2001, ApJ, 549, 745
\bibitem[Sakamoto et al.(2004a)]{sakamoto2003a}
	Sakamoto, T., et al. 2004a, ApJ, 602, 875
\bibitem[Sakamoto et al.(2004b)]{sakamoto2003b}
	Sakamoto, T., et al. 2004b, ApJ, accepted
\bibitem[Sari, Piran \& Halpern(1999)]{sari1999}
	Sari, R., Piran, T., \& Halpern, J. P. 1999, ApJ, 519, L17 
\bibitem[Schmidt(2001)]{schmidt2001}
	Schmidt, M. 2001, ApJ, 552, 36
\bibitem[Soderberg et al.(2003)]{soderberg2003}
	Soderberg, A. M., et al. 2004, ApJ, 606, 994
\bibitem[Vlahakis \& K\"onigl(2001)]{vlahakis2001}
	Vlahakis, N., \& K\"onigl, A. 2001, ApJ, 563, L129
\bibitem[Yamazaki et al.(2002)]{yamazaki2002} 
	Yamazaki, R., Ioka K., \& Nakamura T. 2002, \apj, 571, L31
\bibitem[Yamazaki et al.(2003)]{yamazaki2003} 
	Yamazaki, R., Ioka K., \& Nakamura T. 2003, \apj, 593, 941
\bibitem[Yonutoku et al.(2004)]{yonutoku2004}
	Yonutoku, D., Murakami, T., Nakamura, Yamazaki, R., Inoue, A. K., 
	\& Ioka, K 2004, 609, 935 
\bibitem[Zhang \& M\'esz\'aros(2002)]{zhang2002}
	Zhang, B., \& M\'esz\'aros, P. 2002, \apj, 571, 876 
\bibitem[Zhang \& M\'esz\'aros(2003)]{zhang2003}
	Zhang, B., \& M\'esz\'aros, P. 2003, \apj, 581, 1236
\bibitem[Zhang et al.(2004)]{zhang2004}
	Zhang, B., Dai, X., Lloyd-Ronning, N. M., \& M\'esz\'aros, P.
	2004, 601, L119
\bibitem[Zhang, Woosley \& Heger(2004)]{wzhang2004}
	Zhang, W., Woosley, S. E., \& Heger, A. 2004, \apj, 608, 365



\end{thebibliography}
\end{document}